\documentclass{article}
\usepackage{graphicx} % Required for inserting images
\usepackage[utf8]{inputenc}
\usepackage[T1]{fontenc}
\usepackage{newunicodechar}
\usepackage[affil-it]{authblk}
\usepackage{setspace}
\newunicodechar{≈}{\approx}
\newunicodechar{ν}{\nu}
\usepackage{abstract}
\usepackage{amsmath}
\usepackage{amssymb}
\usepackage{placeins}
\usepackage{color}
\usepackage{bbm}

\usepackage{subcaption}
\usepackage{hyperref}
\usepackage{float}
\newunicodechar{∼}{\sim}
\newunicodechar{₂}{$_2$}
\usepackage[
  backend=biber,
  style=nature,          % biblatex nature style
  sorting=none,
  autocite=superscript,  % use \autocite{} for superscripts (or see \let\cite below)
  url=false, doi=false, eprint=false, isbn=false
]{biblatex}
\AtEveryBibitem{%
  \clearlist{publisher}%    remove Publisher
  \clearlist{location}%     remove Location/Address
  \clearfield{urldate}%
  \clearfield{month}%       remove month like (Nov.)
  \clearlist{language}%     remove "en." etc.
  \clearfield{langid}%      also suppress language tags
  \clearfield{number}%  
  \clearfield{note}
}

\let\cite\supercite

\usepackage{geometry}
\title{Correlation-Driven Ballistic Exciton Expansion 

in a Moiré Bose-Fermi Mixture}

\author[1,{\#}]{Shibin Deng}

\author[1,{\#}]{Jonas M. Peterson}

\author[2]{Maximilian Hauer}

\author[3]{Jonas Reimann}

\author[4]{Heonjoon Park}

\author[3]{Ammon Fischer}

\author[5]{Takashi Taniguchi}

\author[5]{Kenji Watanabe}

\author[4,6]{Xiaodong Xu}

\author[2,3]{Dante M. Kennes}

\author[1]{Libai Huang\textsuperscript{*}}
\affil[1]{Department of Chemistry, Purdue University, West Lafayette, IN 47907, USA}

\affil[2]{Institut für Theorie der Statistischen Physik, RWTH Aachen University, 52056 Aachen, Germany}

\affil[3]{Max Planck Institute for the Structure and Dynamics of Matter, Center for Free Electron Laser Science, 22761 Hamburg, Germany}

\affil[4]{Department of Physics, University of Washington, Seattle, WA 98195, USA}

\affil[5]{Research Center for Electronic and Optical Materials, National Institute for Materials Science, 1-1 Namiki, Tsukuba 305-0044, Japan}

\affil[6]{Department of Materials Science and Engineering, University of Washington, Seattle, WA 98195, USA}

\affil[ ]{\textsuperscript{\#}These authors contributed equally.}

\affil[*]{Corresponding author: libai-huang@purdue.edu}

\date{}

\begin{document}
\maketitle
% (Blank line removed to avoid possible hidden character causing math mode error)
\newpage\section*{Abstract}
Moir\'e superlattices of transition-metal dichalcogenides (TMDs) host strongly interacting Bose--Fermi mixtures in which bosonic excitons coexist with correlated electron lattices. Using ultrafast, time- and energy-resolved photoluminescence (PL) microscopy, we show that strong exciton--electron and exciton--exciton repulsion can enable ballistic exciton transport in a WSe$_2$/WS$_2$ heterobilayer. This ballistic transport is energy selective: repulsive interactions populate higher-energy moiré exciton bands, including the doubly occupied exciton state \(XX\) and the exciton–electron co-localized state \(Xe\), exhibiting mean-squared displacement (MSD) with a quadratic time dependence, \(\mathrm{MSD} \propto t^2\). Transport in the \(Xe\) band is jointly governed by exciton and electron correlations. Exciton correlations suppress relaxation into the lower-energy localized exciton state, while generalized Wigner-crystal (GWC) order at fractional electron fillings enhances Auger recombination and the group velocity of the ballistic expansion. A two-dimensional Bose--Fermi Hubbard model solved using time-dependent mean-field methods qualitatively reproduces the measured exciton transport and time-dependent response. It further confirms that strong cross-species interactions allow the electron crystal to perforate the exciton Mott background, enhancing exciton motion. Our results establish moir\'e TMDs as highly tunable platforms for realizing strongly interacting Bose--Fermi mixtures.
 
\newpage
\section*{Introduction}
Moiré superlattices of transition-metal dichalcogenides (TMDs) have emerged as a solid-state platform to explore strongly correlated Bose-Fermi mixtures, in which long-lived interlayer excitons (bosons) coexist and interact with correlated electron lattices (fermions).\cite{park_dipole_2023,wang_intercell_2023,chen_tuning_2022,arsenault_two-dimensional_2024,mathey_competing_2006,sugawa_interaction_2011,yan_collective_2024} Flat moiré bands localize electrons or holes into lattice sites with strong on-site repulsion, enabling Mott and generalized Wigner-crystal (GWC) states, while the same periodic potential traps long-lived, dipolar excitons with strong exciton-exciton and exciton-charge interactions.\cite{zhang_controlling_2023,xu_correlated_2020,mak_semiconductor_2022,liu_signatures_2021, chen_tuning_2022, regan_mott_2020, gao_excitonic_2024, wang_moire_2021, fan_magnetic_2025, zerba_tuning_2025} Recent experiments have established that excitons are exquisitely sensitive to correlated electronic orders: exciton spectra encode fermionic incompressibility and magnetism; exciton density waves and valley-polarized excitonic insulators can be stabilized by Coulomb-coupled bilayers; and optically injected excitons in a single moiré bilayer exhibit signatures of a bosonic Mott phase with suppressed transport and large interaction gaps.\cite{zeng_exciton_2023,xu_correlated_2020,cai_signatures_2023, xie_long-lived_2024,anderson_trion_2024,wang_hidden_2025} These advances highlight moiré TMDs as controllable Bose–Fermi Hubbard platforms in which excitonic and electronic orders can be achieved by combined electrostatic and optical control.\newline

\noindent Despite rapid progress, the dynamics of coupled bosonic and fermionic orders remains largely unexplored. Dissipative pathways are expected to play an important role for non-equilibrium phase transitions,\cite{ma_dissipatively_2019,dagvadorj_first-order_2021,tomita_observation_2017,Meneghini2025} yet most studies emphasize equilibrium phase diagrams and steady-state spectroscopies. Although excitons have been widely deployed as sensors of electronic phases, using screening-induced shifts, linewidth changes, and intensity modulations, they are typically treated as spectators, \cite{xie_long-lived_2024,wang_hidden_2025} and time-resolved probes of exciton–electron interplay are scarce. Importantly, excitonic coherent dynamics, has not yet been observed in TMD moiré superlattices,\cite{lagoin_key_2021,gotting_moire-bose-hubbard_2022,paik_excitons_2024} underscoring the need for dynamical approaches. In the absence of electrons, unprecedented control of dipolar excitons revealing slow, “frozen” non-equilibrium dynamics of exciton Mott insulators in moiré lattices was demonstrated.\cite{deng_frozen_2025}  However, the hierarchy of timescales governing exciton–exciton versus exciton–charge coupling is unknown, obscuring the microscopic routes by which non-equilibrium phases emerge for mixtures of excitons and electrons. At the mechanism level, the interplay is expected to arise from a competition among on-site exciton–electron repulsion, inter-site interactions, screening, and possible polaronic dressing.\newline

\noindent Here, we address these open questions by time-resolved spectroscopy and imaging of exciton dynamics in a dual-gated WSe$_{\rm{2}}$/WS$_{\rm{2}}$ moiré heterobilayer. We independently controlled the fermionic filling across GWC orders while optically preparing a bosonic exciton Mott insulator. We directly track the coupled dynamics of the two intertwined orders.  Strong exciton–electron repulsion drives excitons into a higher-energy moiré band, where spectral- and time-resolved imaging reveals ballistic transport. This ballistic motion is governed by the correlation of excitons as well as by the  correlation of the electron crystal, highlighting the strong cross-specie interaction in this moiré Bose–Fermi mixture. A two-dimensional Bose–Fermi Hubbard model on the moiré lattice, solved using a time-dependent mean-field approach, reveals the key mechanisms: the electronic GWCs excites the excitons to a higher energy state and perforate the excitonic Mott state. Our results highlight the potential for dynamical engineering in moiré Bose–Fermi mixtures.

\section*{Results }
\begin{figure}[H]
    \centering
    \captionsetup{skip=0pt}
    \vspace{-1em}
    \includegraphics[width=0.8\linewidth]{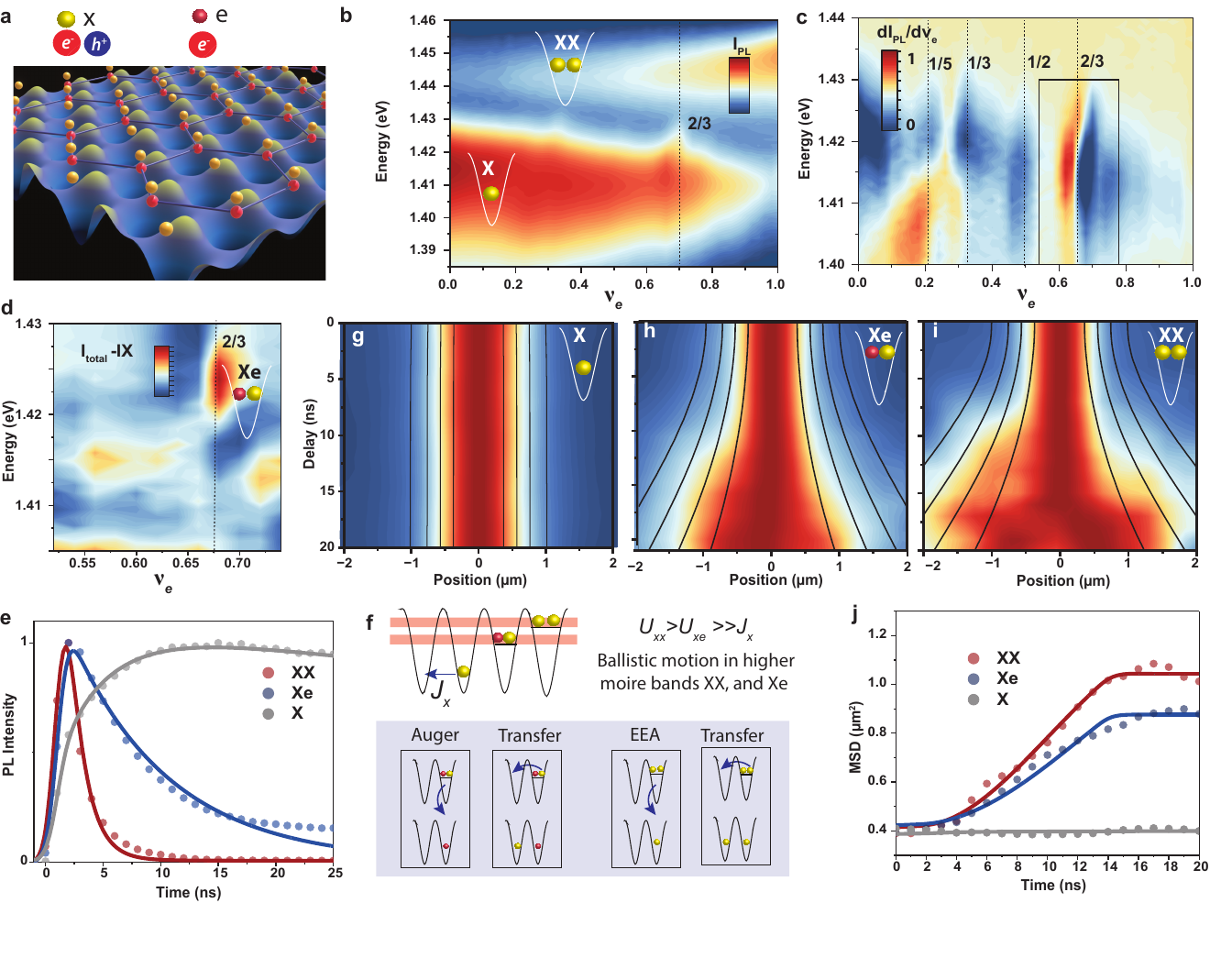}
   \caption{\textbf{Ballistic exciton motion driven by exciton--exciton and exciton--electron repulsion.}
\textbf{a}, Schematic of the coexistence of an electron generalized Wigner crystal (GWC) at electron filling \(\nu_{\rm e}=2/3\) and an exciton Mott insulator at unity exciton filling \((\nu_x=1)\). 
\textbf{b}, Electron-filling-dependent photoluminescence (PL) spectra showing anomalies at GWC fractional fillings, including the emergence of a blue-shifted shoulder. The lowest interlayer exciton band is denoted X, while the higher-energy doubly occupied exciton band is denoted \(XX\). 
\textbf{c}, First derivative of the PL intensity with respect to electron filling, \(\mathrm{d}I_{\rm PL}/\mathrm{d}\nu_e\), highlighting GWC order at \(\nu_e=1/5,\,1/3,\,1/2,\) and \(2/3\). 
\textbf{d}, Subtraction of lowest exciton emission from the total PL intensity near \(\nu_e=2/3\), emphasizing the blue-shifted \(Xe\) shoulder that arises from excitons co-localized with electrons on the same moir\'e sites. 
\textbf{e}, Time-resolved PL decays of the \(XX\), \(Xe\), and IX bands with exponential fits.
\textbf{f}, Schematic of the interaction-induced higher-energy \(XX\) and \(Xe\) bands. The effective interaction hierarchy is \(U_{xx}>U_{xe}\gg J_x\), where \(U_{xx}\) is the exciton--exciton repulsion, \(U_{xe}\) is the exciton--electron repulsion, and \(J_x\) is the nearest-neighbor hopping of the lowest IX band. Both \(XX\) and \(Xe\) are higher-energy moir\'e exciton bands with enhanced effective hopping relative to X. The lower panels illustrate the dominant decay and redistribution pathways, including Auger recombination, exciton transfer, exciton--exciton annihilation (EEA), and transfer from the \(XX\) configuration. 
\textbf{g--i}, Spectrally resolved, time-dependent PL images tracking the spatial expansion of the X, \(Xe\), and \(XX\) bands, respectively. The higher-energy \(Xe\) and \(XX\) bands exhibit much faster spatial expansion than the lowest X band. 
\textbf{j}, Mean-squared displacement (MSD) as a function of delay time. The \(Xe\) and \(XX\) bands show a pronounced quadratic time dependence, \(\mathrm{MSD}(t)\propto t^2\), consistent with ballistic exciton motion, whereas IX shows limited expansion. The larger ballistic velocity of \(XX\) is consistent with its higher emission energy and stronger interaction energy relative to \(Xe\).}
\label{fig:fig1}
\end{figure}

\textbf{Strong correlation between excitons and electrons}. In this work, we investigate the  competing  excitonic and electronic correlations with strong interspecies interactions in the same lattice (schematically illustrated in Fig.~\ref{fig:fig1}a).  To explore this, we study nearly 60° twisted (H-stacked) WS₂/WSe₂ heterobilayers; the detailed device configuration is shown in Extended Data Fig.~1. This combination of materials and the twist angle have previously been shown to host strong correlations of both excitons and charge carriers, including Mott insulators of both charges and excitons at integer fillings as well as generalized Wigner crystals at fractional fillings.\cite{huang_correlated_2021, li_imaging_2021, xu_correlated_2020} In addition, a mixed exciton–charge Mott insulator can form when the combined filling of excitons and charge carriers equals one per moiré unit cell\cite{xiong_correlated_2023}. The moiré superlattice has a lattice constant of $L_M\approx$8 nm and the lowest lying exciton is an interlayer exciton \(X\) at around 1.415 eV (Fig.~\ref{fig:fig1}b). Electrostatic gating controls the electron density, while the exciton density is determined by the photoexcitation intensity using a 740~nm pump resonant with WSe$_{\rm 2}$ (Methods and Extended Data Fig.~1). In this way, the fermion (electron) and boson (exciton) populations are independently tunable, enabling controlled studies of the dynamics and correlations in Bose–Fermi mixtures. In what follows, we focus on electron-exciton interaction (doping electrons). Unless otherwise noted, all optical measurements were performed at 6 K.\newline

\noindent Steady-state PL, plotted versus electron filling \(\nu_e\), exhibits pronounced intensity enhancements at critical fractional fillings \(\nu_c\) (e.g., \(1/5,\,1/3,\,1/2,\) and \(2/3\)), consistent with prior reports (Fig.~\ref{fig:fig1}b).\cite{wang_intercell_2023,regan_mott_2020,xu_correlated_2020} The data in Fig.~\ref{fig:fig1}b were acquired under pulsed excitation for \(0 \le \nu_e \le 1\); complementary CW-excitation spectra over a wider \(\nu_e\) range are shown in Extended Data Fig.~1. The enhancement is commonly ascribed to reduced electron screening and exciton--electron scattering when the electronic system enters correlated insulating states. The pronounced sensitivity of the PL intensity has, in turn, been used as a sensor of GWC order of the electrons. A second higher-energy exciton band, denoted \(XX\), is also observed near 1.44~eV and corresponds to two excitons occupying the same moir\'e site (Fig. ~1b).\cite{deng_frozen_2025} Previous work also showed that moir\'e excitons can bind to charges in neighboring moir\'e cells, producing a redshift of the exciton emission,\cite{wang_intercell_2023} most prominently near \(\nu_e=1/7\). To highlight how the exciton emission evolves with \(\nu_e\), we plot the first derivative of the PL intensity with respect to filling, \(\mathrm{d}I_{\mathrm{PL}}/\mathrm{d}\nu_e\), in Fig.~\ref{fig:fig1}c. Abrupt changes nears fractional fillings that form GWCs. Notably, the PL spectrum develops a clear blue-shifted shoulder alongside the main \(X\) exciton peak at critical fractional fillings, most pronounced at \(\nu_e = 2/3\) (see the zoomed-in view in Fig.~\ref{fig:fig1}d with the \(X\) peak subtracted). We assigned this blue-shifted shoulder to the exciton--electron co-localized band, denoted \(Xe\).\\

\noindent The blue-shifted \(Xe\) emission at 1.425~eV is consistent with strong on-site exciton--electron repulsion, \(U_{Xe}\). When electrons crystallize at GWC fillings, the probability that an exciton becomes co-localized with an electron on the same moir\'e site increases, promoting the exciton into a higher-energy moir\'e band at the GWC sites. We note that this shift is smaller than the blue shift observed at \(\nu_e=1\), where electrons form a Mott insulator;\cite{wang_intercell_2023, park_dipole_2023} this difference is consistent with weaker screening in the GWC state compared with the electron Mott state. These data in Fig.~\ref{fig:fig1} were acquired at an initial exciton density of \(N_{0}\approx0.3\), corresponding to an excitation power of \(33~\mathrm{nW}\), where \(N_{0}\) is the average number of excitons per moir\'e site.\\ 

\noindent The \(Xe\) shoulder lies at lower energy than the \(XX\) emission arising from two excitons occupying the same moiré site (Fig.~\ref{fig:fig1}b), consistent with an effective interaction hierarchy \(U_{XX} > U_{Xe}\), where \(U_{XX}\) and \(U_{Xe}\) denote the exciton--exciton and exciton--electron repulsion energies, respectively. Importantly, both interaction energies greatly exceed the nearest-neighbor hopping \(J_X\) of the single-exciton \(X\) band, which has been reported to be less than \(1~\mathrm{meV}\).\cite{deng_frozen_2025} Time-resolved PL measurements show that the \(Xe\) shoulder at \(\nu_e = 2/3\) decays substantially faster than the lower-energy \(X\) peak at 1.415~eV associated with singly occupied moir\'e sites (Fig.~\ref{fig:fig1}e). The lifetime of the \(Xe\) population is limited by two competing processes (Fig.~\ref{fig:fig1}f). First, exciton--electron co-localization enhances three-particle Auger recombination, in which exciton recombination non-radiatively transfers energy to the  electron, leading to a faster decay.\cite{haug_auger_1983, wang_auger_2006} Second, the exciton can transfer from the electron-occupied site to a neighboring empty moir\'e site, which removes population from the \(Xe\) configuration without requiring recombination. Therefore, the measured \(Xe\) decay reflects both Auger recombination and redistribution of excitons out of the co-localized state. The \(XX\) population is more rapidly depleted by exciton--exciton annihilation (EEA) or by redisttribution of excitons, as shown by the red curve in Fig.~\ref{fig:fig1}e), with a characteristic timescale of \(\tau_{XX}\sim 3\)~ns, as discussed in our previous work.\cite{deng_frozen_2025} The dynamics processes are illustrated in Fig.~1f.\\

\noindent\textbf{Ballistic Exciton Transport of Higher-Energy Moiré Band}. Because both \(XX\) and \(Xe\) are higher-energy moir\'e exciton bands, they are expected to have larger dispersion and enhanced effective hopping compared with \(J_X\).\cite{brem_bosonic_2023}  To investigate this, we perform spectrally resolved, time-dependent imaging to track the expansion of the exciton cloud in \(X\), \(XX\) and \(Xe\) moir\'e bands (Fig.~ 1g-j). A diffraction-limited femtosecond pulse initializes a Gaussian exciton distribution, and the spatially and spectrally resolved PL is recorded by raster-scanning a single-photon counting detector in the image plane (Methods; Extended Data Fig. 2). A similar method has been extensively applied to image exciton diffusion.\cite{akselrod_visualization_2014,kulig_exciton_2018, ginsberg_spatially_2020}.\\

\noindent We first investigate the transport dynamics at a low initial exciton density of \(N_{0}\approx 0.3\), which is below the threshold for forming an exciton Mott state,\cite{deng_frozen_2025} where correlation effects are expected to be weak. At this density, many moir\'e sites remain empty, although pulsed excitation still generates small populations of the \(XX\) and \(Xe\) states. Energy-resolved transport measurements reveal substantially faster exciton motion in the two higher-energy moiré exciton bands, \(XX\) and \(Xe\), than in the lowest-energy \(X\) band (Fig.~\ref{fig:fig1}g--i). To quantify the transport dynamics, we compute the mean-squared displacement (MSD, Fig.~\ref{fig:fig1}j) from the time-dependent exciton probabilities \(P_i(t)\) on moir\'e sites \(i\)(Fig.~\ref{fig:fig1}g--i),
\begin{equation}
\mathrm{MSD}(t)=\sum_{i} P_i(t)\,\bigl\lVert \mathbf{r}_i-\mathbf{r}_0 \bigr\rVert^{2}, \label{eq:MSD}
\end{equation}
where \(\mathbf{r}_0\) is the initial excitation position.

\noindent Most notably, both the \(XX\) and \(Xe\) bands exhibit a quadratic time dependence, \(\mathrm{MSD}(t)\propto t^{2}\), over the first 10 ns (Fig.~\ref{fig:fig1}g-i), whereas the \(X\) state shows only limited spatial expansion. This quadratic time dependence is a hallmark of ballistic transport, in contrast to diffusive motion, for which \(\mathrm{MSD}(t)\propto t\). Furthermore, at later times, the spatial profile deviates substantially from a Gaussian distribution: the initial central peak evolves into two spatially separated peaks that propagate outward (Extended Data Fig.3). Such peak splitting is also a hallmark of ballistic transport, although a broad distribution of group velocities or a limited ballistic propagation distance can broaden the peaks and partially obscure this feature (Extended data Fig. 4-6). The observed \(\mathrm{MSD}(t)\propto t^{2}\) behavior also cannot be explained by nonlinear population dynamics such as exciton--exciton annihilation (EEA) or Auger recombination, as detailed in Extended Data Fig.~5. These loss processes deplete the exciton population more rapidly at the excitation center and can therefore artificially broaden the apparent spatial distribution. However, such effects produce a sublinear time dependence of the MSD, rather than the ballistic scaling observed here.\\

\noindent To further analyze the spatiotemporal dynamics, we developed a kinetic model in which the higher-energy \(XX\) and \(Xe\) bands are dynamically coupled to the lowest \(X\) band, as shown in Extended Data Fig.~4. This model explicitly accounts for both transport and population dynamics among the different moir\'e exciton bands. 
The essential features of the model are summarized below. The full model incorporates a distribution of outward velocities, as described in Methods.

\[
\frac{\partial U}{\partial t}
=
P(r,t)
-
k_UA(L)U
-
k_{\mathrm{loss}}U
-
\frac{1}{r}\frac{\partial}{\partial r}\left(rv_gU\right)
\]

\[
\frac{\partial L}{\partial t}
=
k_UA(L)U
-
\gamma_LL
+
D_L\frac{1}{r}\frac{\partial}{\partial r}
\left(r\frac{\partial L}{\partial r}\right)
\]

Here, $U(r,t)$ and $L(r,t)$ denote the upper- and lower-energy exciton band areal densities, respectively, and $v_g$ is the outward propagation velocity of the upper band. The excitation profile is given by $P(r,t)$. The transfer rate $k_U$ describes relaxation from the upper band into the lower band, while the availability function $A(L)$ accounts for the reduced transfer probability as lower moiré band being occupied. The parameter $k_{\mathrm{loss}}$ represents competing upper-state loss that does not populate the lower-energy state. Finally, $\gamma_L$ and $D_L$ denote the lower-energy state decay rate and diffusivity, respectively.

%{\color{blue}
%The essential features of the model are summarized by the following coupled transport equations.

%\[
%\frac{\partial u}{\partial t}
%=
%f(v)P(r,t)
%-
%k_UA(L)u
%-
%k_{\mathrm{loss}}u
%-
%\frac{1}{r}\frac{\partial}{\partial r}\left(rvu\right)
%\]

%\[
%\frac{\partial L}{\partial t}
%=
%k_UA(L)U
%-
%\gamma_LL
%+
%D_L\frac{1}{r}\frac{\partial}{\partial r}
%\left(r\frac{\partial L}{\partial r}\right)
%\]

%Here, $u(r,v,t)$ denotes the upper-state exciton population per unit outward velocity $v$, while $L$ is the lower-energy moiré exciton population. The normalized distribution $f(v)$ describes the positive velocities of excitons entering the upper state, and the total upper-state population is $U=\int_0^\infty u\,dv$. The distribution is Gaussian, with center parameter $v_g$ and width $\sigma_v$; $v_g$ is extracted from the fit. The excitation profile is given by $P(r,t)$. The transfer rate $k_U$ describes relaxation from the mobile exciton population into the lower energy state, while the availability function $A(L)$ accounts for the reduced transfer probability as the lower energy state becomes occupied. The parameter $k_{\mathrm{loss}}$ represents competing upper-state loss that does not populate the lower-energy state. Finally, $\gamma_L$ and $D_L$ denote the lower-energy moiré exciton decay rate and diffusivity, respectively.
%}

The model allows us to disentangle changes in the MSD arising from population dynamics from those due to spatial transport and to extract ballistic group velocities, \(V_g\), of approximately \(0.04~\mu\mathrm{m/ns}\) and \(0.03~\mu\mathrm{m/ns}\) for the \(XX\) and \(Xe\) bands, respectively. The fits are shown as black lines on top of the transport imaging in Fig.~\ref{fig:fig1}g-i) as well in Fig. ~1e and Fig.~1j. The larger ballistic velocity observed for \(XX\) is consistent with its higher energy and stronger interaction energy relative to \(Xe\), which leads to a more dispersive exciton band. We note that ballistic transport could also occur at exciton filling below unity in the lowest moir\'e exciton band in the absence of disorder. In TMD heterobilayers, however, unavoidable disorder and scattering typically leave the effective exciton hopping $J_X$ smaller than both the thermal energy $k_{\mathrm B}T$ and the site-to-site energetic disorder $\Delta$.\cite{lagoin_key_2021,gotting_moire-bose-hubbard_2022,pollet_absence_2009,dupuis_superfluidbose-glass_2024} As a result, ballistic transport has not been observed for lowest-band moir\'e excitons. Consistent with this, previous WSe$_2$/WS$_2$ measurements at charge neutrality observed only diffusive motion below unity filling.\cite{deng_frozen_2025,gao_excitonic_2024}\newline

\noindent\textbf{GWC-enhanced ballistic exciton transport} We next investigate how electronic and exciton orders modifies the dynamics and transport of \(Xe\) by sweeping the electron filling \(\nu_e\) across the GWC fillings (exciton transport PL images are shown in Extended Data Figs.~7 and 8). To separate dynamics related to exciton correlations from single-exciton behavior, we compare a high-density regime, where excitons can form a Mott-insulator state with an initial exciton density of \(N_{0}\approx 5\), with a low-density regime, where excitons remain weakly correlated with \(N_{0}\approx 0.3\).  %Note that \(N_0\) is distinct from the exciton filling factor \(\nu_{\rm ex}\): \(N_0\) denotes the estimated initial injected exciton density at the center of the Gaussian pump profile, normalized by the moiré site density.
Note that $N_0$ is distinct from the exciton filling factor $\nu_{\mathrm{ex}}$:
$N_0$ denotes the spatially averaged initial injected exciton density, normalized
by the moiré site density. After pulsed excitation, excitons rapidly expand spatially so \(\nu_{\rm ex}\) evolves in both space and time; as shown in our previous report,\cite{deng_frozen_2025} for \(N_0 > 1\), the injected exciton density is sufficiently high for the evolving exciton ensemble to enter the Mott-insulator regime during its subsequent relaxation and expansion dynamics. The PL intensity from \(XX\) increases with exciton density, as higher exciton filling increases the probability of two excitons occupying the same moiré site (Extended Data Fig. 1).\\

\begin{figure}[H]
    \centering
    \includegraphics[width=1\linewidth]{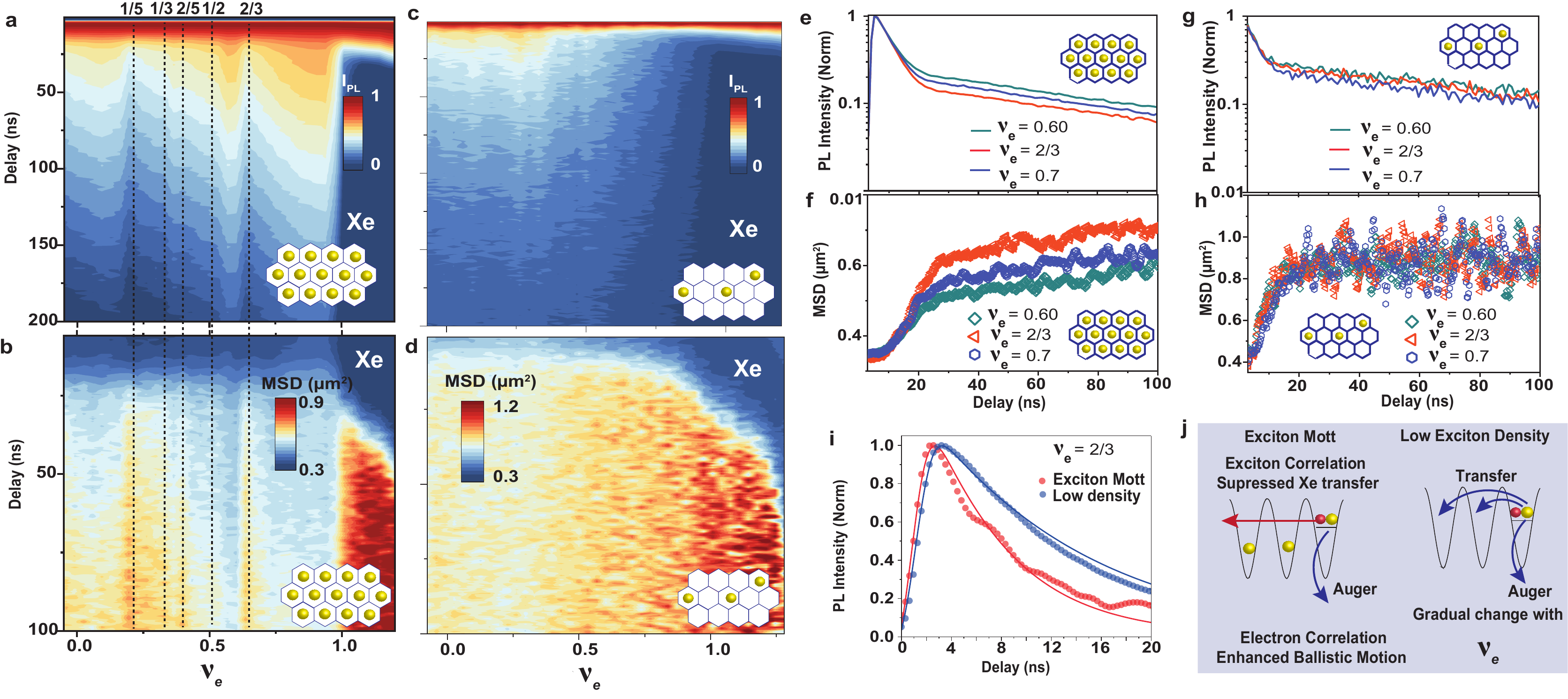}
    \caption{\textbf{GWC-enhanced ballistic exciton transport due to Interplay between Exciton and Electron correlations.}
\textbf{a,b,} Time- and electron-filling-dependent PL intensity (\textbf{a}) and mean-squared displacement (MSD, \textbf{b}) of $Xe$ in the exciton Mott regime. Dashed lines mark representative fractional electron fillings.
\textbf{c,d,} Corresponding PL intensity (\textbf{c}) and MSD (\textbf{d}) measured in the low-exciton-density regime. In contrast to the exciton Mott regime, the filling dependence of the $Xe$ dynamics and transport is substantially weaker at low exciton density.
\textbf{e--h,} $Xe$ dynamics and transport near the critical filling $\nu_e=2/3$. Time-dependent PL intensity (\textbf{e,g}) and MSD (\textbf{f,h}) at $\nu_e=0.59$, $2/3$, and $0.71$ for the exciton Mott (\textbf{e,f}) and low-exciton-density (\textbf{g,h}) regimes. In the exciton Mott regime, $Xe$ decay is faster and ballistic transport is enhanced at $\nu_e=2/3$, whereas only a gradual variation with $\nu_e$ is observed at low exciton density.
\textbf{i,} Direct comparison of the normalized $Xe$ PL dynamics at $\nu_e=2/3$ in the exciton Mott and low-exciton-density regimes.
\textbf{j,} Schematic illustrating the interplay between exciton and electron correlations. In the exciton Mott regime, exciton correlations suppress $Xe$ transfer between moir\'e sites, enhancing the sensitivity of $Xe$ dynamics to the underlying electron correlations. At the generalized Wigner crystal (GWC) filling, electron correlations enhance recombination, shortening the $Xe$ lifetime while enhancing the ballistic transport. At low exciton density, $Xe$ transfer remains efficient and the dynamics vary only gradually with electron filling.}
\label{fig:fig2}
\end{figure}

\noindent In the high-density exciton-Mott regime, tuning \(\nu_e\) through the GWC fillings produces abrupt changes in the \(Xe\) lifetime (Fig.~\ref{fig:fig2}a), with pronounced features at \(\nu_e=1/5\), \(1/3\), \(2/5\), \(1/2\), and \(2/3\). The largest deviations occur within the first \(\sim 30\)~ns, when exciton--electron Auger recombination is most effective. These rapid lifetime changes indicate that Auger recombination is enhanced when the electrons form spatially ordered GWC states. The corresponding MSD also exhibits sharp enhancements at the same fractional fillings (Fig.~\ref{fig:fig2}b), demonstrating that the shortened lifetime is accompanied by faster spatial expansion of the \(Xe\) exciton cloud. The simultaneous shortening of the \(Xe\) lifetime and enhancement of exciton transport indicate strong cross-species correlations. The strongest response occurs near \(\nu_e=2/3\), consistent with the PL-intensity anomalies shown in Fig.~\ref{fig:fig1}c,d. In stark contrast, in the low-density regime, where excitons do not form a Mott state, the exciton dynamics evolve smoothly with \(\nu_e\) (Fig.~\ref{fig:fig2}c). Increasing \(\nu_e\) gradually increases the probability of exciton--electron co-localization, leading to a monotonic reduction of the \(Xe\) lifetime without the abrupt features observed in the Mott regime (Fig.~\ref{fig:fig2}c). The corresponding transport also increases smoothly with \(\nu_e\) (Fig.~\ref{fig:fig2}d), consistent with a gradual enhancement of intersite hopping. The contrasting dynamics and transport behaviors of the exciton Mott and low-density regimes can be clearly seen when we zoom in on fillings near the critical value, $\nu_e=\frac{2}{3}$ (Fig.~\ref{fig:fig2}e--h).\\

\noindent\textbf{Exciton and Electron Correlations Modulate Ballistic Excitons.} The strong enhancement of the \(Xe\) response at GWC fillings in the exciton Mott regime indicates that the dynamics of the \(Xe\) band are governed by the interplay between exciton and electron correlations. Interestingly, the \(Xe\) decay is substantially slower in the high-density exciton Mott regime (Fig.~\ref{fig:fig2}i). In the exciton Mott state, near-unity occupation of the lowest \(X\) band leaves few vacant moiré sites available for an exciton in the \(Xe\) configuration to transfer into. The \(Xe \rightarrow X\) transfer pathway is therefore blocked, suppressing depletion of the \(Xe\) population and extending its lifetime (Fig.~\ref{fig:fig2}i). At the same time, when electron GWC order is present, the correlated spatial arrangement of excitons and electrons enhances their co-localization and thereby increases the probability of Auger recombination. The competition between Mott-induced blockade of \(Xe \rightarrow X\) relaxation and GWC-enhanced Auger recombination makes the \(Xe\) dynamics particularly sensitive to electron crystallization in the exciton Mott regime, whereas this sensitivity is much weaker at low exciton density (illustrated in Fig.~\ref{fig:fig2}j). The measured exciton-density-dependent dynamics are well reproduced by the kinetic model incorporating blockade of the \(Xe \rightarrow X\) relaxation pathway (Fig.~\ref{fig:fig2}i).

\begin{figure}[H]
     \centering
     \includegraphics[width=1\linewidth]{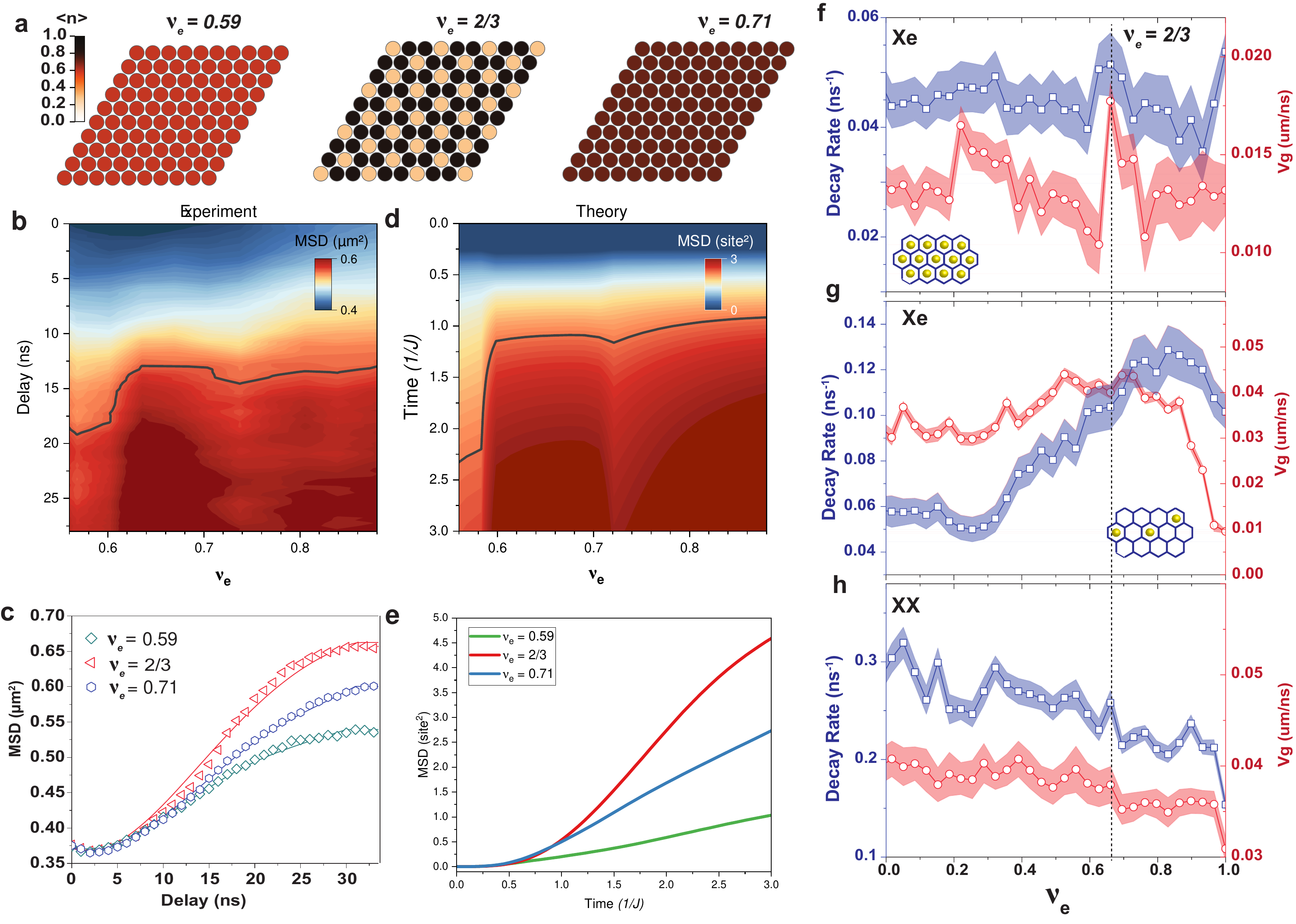}
\caption{\textbf{Simulations of a two-dimensional Bose-Fermi Mixture.}
\textbf{a,} Calculated electron density $\langle n\rangle$ in the moir\'e lattice at representative electron fillings $\nu_e=0.59$, $2/3$, and $0.71$, showing the formation of the generalized Wigner crystal (GWC) at $\nu_e=2/3$.
\textbf{b,} Experimental MSD of $Xe$ as a function of delay time and electron filling $\nu_e$ in the exciton Mott regime.
\textbf{c,} Experimental $Xe$ MSD dynamics at $\nu_e=0.59$, $2/3$, and $0.71$, showing enhanced ballistic transport at the GWC filling $\nu_e=2/3$.
\textbf{d,} Calculated MSD as a function of time and electron filling, reproducing the enhanced exciton transport near $\nu_e=2/3$ observed experimentally.
\textbf{e,} Calculated MSD dynamics at $\nu_e=0.59$, $2/3$, and $0.71$, showing the enhanced propagation at $\nu_e=2/3$.
\textbf{f--h,} Filling-dependent decay rate (blue, left axis) and ballistic velocity $V_g$ (red, right axis) for $Xe$ in the exciton Mott regime (\textbf{f}), $Xe$ in the low-exciton-density regime (\textbf{g}), and $XX$ (\textbf{h}). Shaded regions represent uncertainties. The vertical dashed line marks $\nu_e=2/3$. In the exciton Mott regime, $Xe$ exhibits a pronounced enhancement in both decay rate and ballistic velocity at $\nu_e=2/3$, whereas the corresponding quantities evolve more gradually with filling at low exciton density. The $XX$ state shows no sensitivity to electron correlations.}
\label{fig:fig3}
 \end{figure}

\noindent\textbf{Simulations of a Two-Dimensional Bose-Fermi Mixture.} To connect to the underlying microscopic mechanism of exciton transport in the presence of both exciton and electron correlations, we perform time-dependent numerical simulations using a two-dimensional extended Bose Hubbard model on a triangular lattice, solved by the Keldysh formalism using a Hartree-Fock mean-field approximation\cite{Stefanucci,SchlünzenRev}. Moiré excitons feature long-ranged repulsive interaction due to their dipolar nature. To simplify, we model them as bosons interacting via a repulsive interaction term $\propto \frac{V^{xx}}{|\mathbf{r}_i-\mathbf{r}_j|^3}$ where $V^{xx}$ is the strength of the dipolar interaction between an exciton at $\mathbf{r}_i$ and at $\mathbf{r}_j$. We explicitly introduce two Moiré-bands to model the energetic shift between $X$ and $Xe$ observed in the experiment. Because these two bands have different kinetics, the upper band is chosen to be much more dispersive. To focus on electron-induced effects, a third band corresponding to $XX$ is omitted. The electrons are modeled as spinless fermions with a nearest-neighbor interaction $V^{ee}$. Electronic filling values around critical filling $\nu_e=\frac{2}{3}$ are examined and a phase transition between the fluid- and GWC phase is observed, and the initial electron densities are plotted in (Fig.~\ref{fig:fig3}a). The finite lifetime of excitons is taken into account by an open system approach with Markovian reservoirs. The excitons are prepared in an initial uncorrelated state, where sites in a circular center region are singly occupied in the lower band. Electrons are prepared in an equilibrium initial state, which is computed in the absence of excitons. We then simulate the time evolution of the interacting hybrid system. For more details, see the Method section.\\

\noindent We observe a pronounced enhancement of the MSD in the $Xe$ band at electron filling factors where the electronic density crystallizes. We focus on the region near $\nu_e=\frac{2}{3}$, where the experimentally measured MSD is shown in Fig.~\ref{fig:fig3}b-c and the corresponding simulation results are shown in Fig.~\ref{fig:fig3}d-e. The simulations reproduce key experimental observations, including the abrupt enhancement of transport near the GWC phase and the emergence of ballistic MSD. Importantly, the transport enhancement associated with the GWC phase is significantly greater than the overall trend of increasing MSD with increasing $\nu_e$. This enhancement arises from the fact that GWC increases exciton transfer from the lower band to the more dispersive upper band, followed by rapid transport within the upper band. \newline

\noindent Figure~3f-h summarizes the extracted \(\nu_e\)-dependence on the \(Xe\) group velocity and decay rate. At the high density exciton Mott regime, the \(Xe\) ballistic group velocity show pronounced anomalies at the GWC fillings, most clearly near \(\nu_e=2/3\). In contrast, at low exciton density when an exciton Mott is not established, the \(Xe\) response evolves more smoothly with \(\nu_e\) (Fig.~\ref{fig:fig3}g). The absence of sharp anomalies at low exciton density shows that the features near GWC fillings cannot be explained by electron correlation alone. The extracted \(Xe\) group velocities in the exciton Mott regime are lower than those in the low-density regime, which may arise from density-dependent renormalization of the effective exciton dispersion by many-body exciton--exciton and exciton--electron interactions.\cite{lin_exciton-driven_2022}. In both the Mott and low-density regimes, exciton transport is enhanced near \(\nu_e=1\), where electrons form a Mott insulator. In this case, the unity filling of  electron can reduce the effective moir\'e potential experienced by excitons and thereby facilitate intersite hopping, consistent with recent steady-state diffusion length measurement results.\cite{upadhyay_giant_2025,yan_anomalously_2025}\\

\noindent As a control, we examine the \(XX\) band, which originates from two excitons occupying the same moir\'e site and is therefore governed primarily by exciton--exciton correlations rather than exciton--electron correlations (Fig.~\ref{fig:fig3}h). Unlike \(Xe\), $XX$ shows no corresponding anomaly at the GWC fillings. Thus, neither ballistic transport nor strong exciton correlations alone is sufficient to produce the observed filling-dependent response; the sharp $Xe$ anomalies specifically require coupling between the correlated exciton and electron lattices.\\

\noindent We emphasize that the agreement between experiment and simulation should be viewed as a qualitative one. Although two exciton bands are included in the simulations, this simplified model cannot fully resolve the complex excitonic energetics present in the experiment. Moreover, we have checked that for the simulation parameter shown here the electronic background can be treated as a static where the back action of the exciton dynamics on the electrons can be disregarded.  This electronic background enters the simulation only through its density within the mean-field approximation; electronic correlations and fluctuations are therefore not explicitly included. Despite these simplifications, the calculations consistently capture the central mechanism: electron crystallization enhances the promotion of excitons into the higher, more dispersive states, thereby increasing exciton transport. Together, theory and experiment indicate that strong exciton--electron correlations, rather than screening alone, are the primary driver of the observed gate-dependent exciton decay and group velocity.\\

\begin{figure}[H]
    \centering
    \includegraphics[width=1\linewidth]{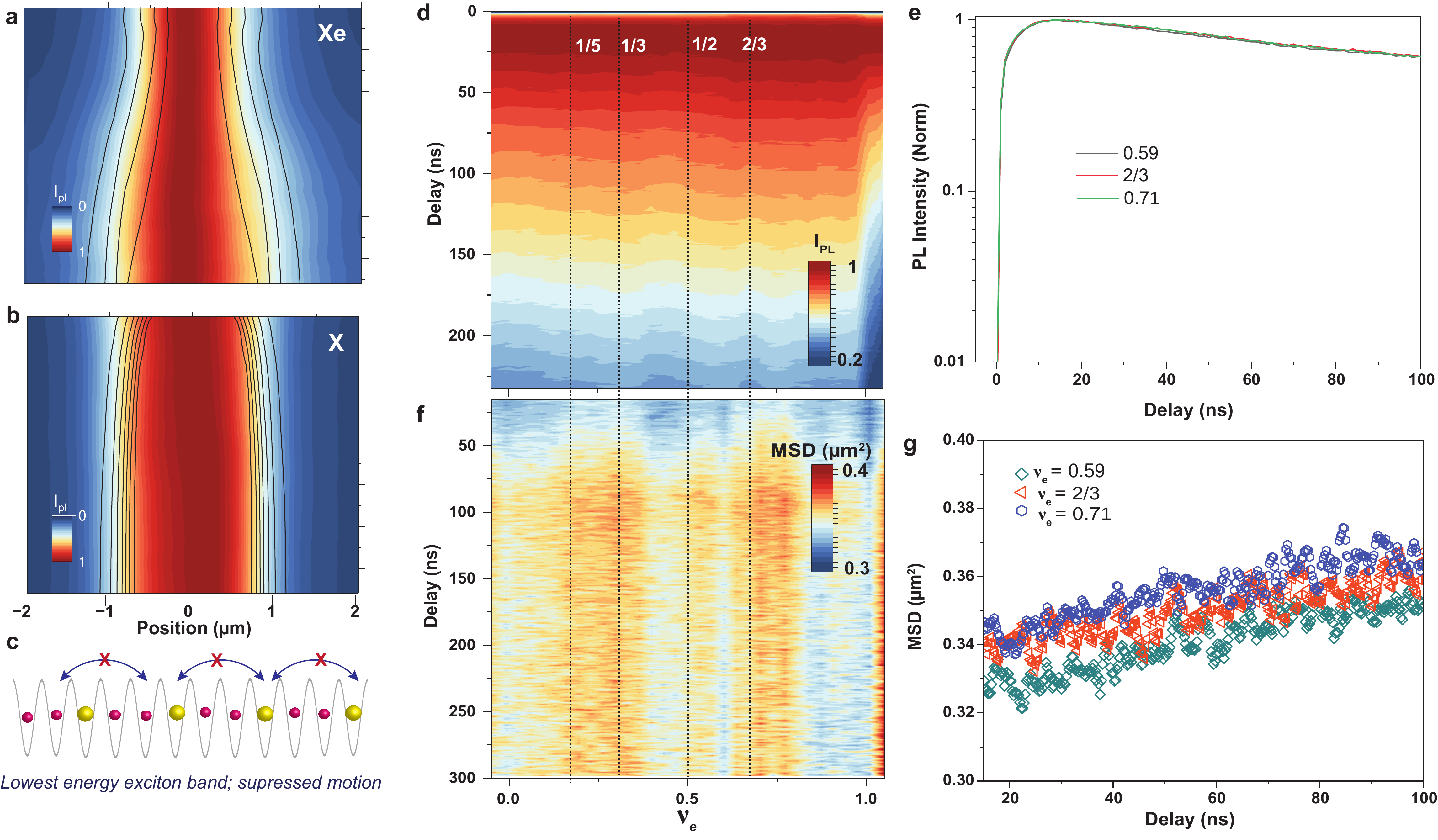}
    \caption{\textbf {Exciton dynamics and transport in the lowest moiré band.} 
    \textbf{a-b} Coupled spatiotemporal evolution of the higher-energy \(Xe\) band and the lower-energy IX/remaining exciton band, fitted using the kinetic model. The model captures the transfer, decay, and spatial propagation channels that connect the mobile \(Xe\) population to the longer-lived exciton population. 
\textbf{c}, Schematic of a mixed electron--exciton Mott state.
\textbf{d-g}  exciton dynamics and transport in the lowest moiré band probed at 1.415 eV. \textbf{d} False-colour map of time-resolved PL intensity at 1.415~eV (lowest moiré exciton band) in the high-density regime; the emission arises from singly occupied sites and varies only weakly with electron filling. \textbf{e}, Representative PL transients for $\nu_e=0.59,\,2/3,$ and $0.71$, highlighting the weak $\nu_e$-dependence of the lifetime for the lowest band. \textbf{f}, Mean-squared displacement (MSD) versus delay and electron filling $\nu_e$, showing strongly suppressed motion once the mixed Mott state is established. \textbf{g}, MSD for the same fillings, confirming slow expansion and the absence of a ballistic window.} 
\label{fig:fig4}
\end{figure}

\noindent\textbf{Transition to the Mixed Exciton--Electron Mott Insulator}. Having established how exciton and electron correlations regulate the lifetime and transport of \(Xe\), we next investigate its relaxation into the lowest-energy \(X\) band. The coupled spatiotemporal evolution of the higher-energy \(Xe\) band and the lower-energy \(X\) band further clarifies the relaxation pathway during this transition (Fig.~\ref{fig:fig4}a,b). The measured \(Xe\) and \(X\) transport cannot be analyzed as independent channels, because the two populations are dynamically coupled: \(Xe\) can decay through Auger recombination or transfer into lower-energy \(X\) configurations, but if the site is also occupied such channel will be blockaded. This sequence, transient ballistic expansion and subsequent localization into the mixed electron--exciton Mott configuration, provides a time-domain picture of a non-equilibrium mixed-Mott transition. In this longer time scale, exciton motion is strongly suppressed ((Fig.~\ref{fig:fig4}c). Both exciton and electron correlations provide control knobs for tuning the pathway and timescale of the non-equilibrium phase transition.

The transport in the mixed exciton--electron Mott insulator is monitored by imaging the lowest moiré exciton band \(X\) at 1.415~eV (Fig.~\ref{fig:fig4}d-g). The corresponding lifetime is long, $\sim\!150$~ns, and depends only weakly on electron filling $\nu_e$ (Fig.~\ref{fig:fig4}d), as illustrated by representative traces near $\nu_e=2/3$ (Fig.~\ref{fig:fig4}e). This contrasts sharply with the fast decay and the strong $\nu_e$ sensitivity of the Auger recombination of the higher-energy co-localized (electron--exciton) band shown in Fig.~\ref{fig:fig2}a. The $\nu_e$-dependent MSD (Fig.~\ref{fig:fig4}f,g) confirms that, once the mixed Mott state is established, exciton motion is strongly suppressed: the effective diffusion constant is over one order of magnitude smaller than the peak ballistic value $D_{\max}$. The filling-dependent offset in the lowest-band MSD near the fractional fillings (Fig.~4f,g), despite no discernible change in its subsequent slow diffusion, reflects the coupled dynamics of the two states: the enhanced ballistic redistribution of the higher-energy $Xe$ population during the first $\sim 30$~ns is inherited by the lower-energy $X$ population as $Xe$ relaxes into $X$.

\section*{Discussion}
We demonstrate ballistic exciton transport in the higher-energy moiré bands of a WS\(_2\)/WSe\(_2\) moiré heterobilayer. This ballistic transport channel, observed for both the \(XX\) and \(Xe\) bands, is revealed only through combined time-, energy-, and spatially resolved imaging. Importantly, these results establish that cross-species correlations provide a powerful means of controlling exciton dynamics and transport in moiré Bose--Fermi mixtures. A key observation is that ballistic propagation in the higher-energy \(Xe\) band is sustained by two cooperative many-body effects: (i) strong exciton correlations and (ii) electron crystallization. The two-dimensional Bose--Fermi model qualitatively shows that formation of the electron crystal enhances transfer of excitons from the lower X band into the more dispersive Xe band, which promoting rapid exciton motion. In the exciton Mott regime, unity occupation of the lowest exciton band leaves no empty moiré sites into which the exciton in the \(Xe\) configuration can transfer, thereby blocking the \(Xe \rightarrow X\) relaxation pathway and extending the lifetime of the mobile \(Xe\) population. The interplay between exciton and charge correlations accounts for the strong transport enhancements at fractional GWC fillings in the high-density exciton Mott regime and the absence of sharp filling-dependent features in the weakly exciton-correlated regime.\newline
 
\noindent These results provide a framework for engineering interaction-driven phases through tailored coupling between bosonic and fermionic degrees of freedom in moiré superlattices. More broadly, they illustrate how one quantum order can serve as a control knob for another, pointing to time-domain materials design in which phases are programmed by transient interactions and dissipation. For example, engineered dissipation is showed to delay Mott melting and suppress coherence growth in cold-atom Bose–Hubbard lattices via quantum-Zeno suppression of tunneling.\cite{tomita_observation_2017} Our results realize an electronic-order–driven variant of this idea in a solid, where an electron crystal can be used as a control for dissipative pathways of the the exciton order.

\section*{Methods}

\noindent\textbf{Sample assembly and characterization}

Bulk WS₂ and WSe₂ were mechanically exfoliated onto Si substrates capped with 285 nm SiO₂. Monolayer flakes were first screened by optical contrast and then confirmed by atomic force microscopy (AFM). Crystal axes were identified via second-harmonic generation (SHG). Further sample details can be found also in our previous publication \cite{deng_frozen_2025}.

Bottom gates were fabricated by evaporating Pt contacts onto a preassembled graphite/hexagonal boron nitride (hBN) stack made with a polycarbonate (PC) dry-transfer workflow. Prior to placing the aligned WS₂/WSe₂ bilayer, the gate surface was “polished” using contact-mode AFM (Bruker Dimension Icon). After the heterobilayer transfer, we performed an additional surface clean and used piezoresponse force microscopy (PFM) to extract the moiré wavelength and deduce the relative twist angle. A top graphite/hBN laminate was then transferred to complete the dual-gated device.\newline

\noindent\textbf{Steady-state photoluminescence spectroscopy}

Steady-state (time-averaged) photoluminescence (PL) spectra were acquired with the sample mounted in a temperature-stabilized closed-cycle cryostat (Montana Instruments). Unless otherwise noted, measurements were performed at \(6~\mathrm{K}\). Excitation was provided by optical parametric amplifiers (Light Conversion; \(\sim\!300~\mathrm{fs}\), \(750~\mathrm{kHz}\) repetition rate). A \(100\times\) Zeiss objective (NA \(=0.9\)) positioned inside the cryostat focused \(740~\mathrm{nm}\) pulses to a diffraction-limited spot and collected the epi-emitted PL. For each electron filling, the emission was directed to an Andor imaging spectrometer and dispersed onto a CCD for detection.\newline

\noindent\textbf{Time-resolved PL dynamics and exciton transport imaging}

Ultrafast measurements were performed with the sample in a temperature-stabilized cryostat (Montana Instruments). Unless noted, data were taken at 6 K. Excitation pulses came from optical parametric amplifiers (Light Conversion; $\sim$300 fs, 750 kHz). A 100× Zeiss objective (NA = 0.9) inside the cryostat focused the 740 nm pulses to a diffraction-limited spot and collected epi-emitted PL. The PL was focused by a 300mm achromat to the image plane, where a multimode-fiber tip mounted on a motorized stage sampled the signal and delivered photons to a single-photon avalanche detector (Excelitas SPCM-NIR). Time-correlated single-photon counting (PicoQuant PicoHarp 300) yielded the PL decay curves. For spectrally and spatially resolved PL microscopy, a transmission grating was inserted upstream of the 300 mm lens to disperse the emission before fiber coupling.\newline

\noindent\textbf{Determination of exciton density}

We estimate the injected exciton density per pulse from the pump fluence and sample absorptance.
For a Gaussian pump with fluence profile
\begin{equation}
F(x,y) = F_0 \exp\!\left[-\frac{x^2 + y^2}{2\sigma^2}\right],
\end{equation}
the pulse energy is the integral over the focal plane,
\begin{equation}
\label{eq:pulse_energy}
E \;=\; \iint F(x,y)\,dx\,dy .
\end{equation}
Here, $F_0$ is a normalization constant and $\sigma$ is the Gaussian width.

For a repetition rate $f=750~\mathrm{kHz}$ and average power $P=110~\mathrm{nW}$,
the pulse energy is
\begin{equation}
E \;=\; \frac{P}{f} \;=\; 1.47\times 10^{-13}\ \mathrm{J}.
\end{equation}
With $\sigma = 0.22~\mu\mathrm{m}$, the spatially averaged fluence is
\begin{equation}
\bar{F} \;=\; \frac{E}{4\pi\sigma^2}
\;\approx\; 2.4\times 10^{1}\ \mu\mathrm{J\,cm^{-2}},
\end{equation}
evaluated at the pump photon energy $E_\gamma=1.68~\mathrm{eV}$.
The corresponding spatially averaged photon areal density is
\begin{equation}
N_\gamma \;=\; \frac{\bar{F}}{E_\gamma}
\;\approx\; \frac{24~\mu\mathrm{J\,cm^{-2}}}{1.68~\mathrm{eV}}
\;\approx\; 9\times 10^{13}\ \mathrm{cm^{-2}\,pulse^{-1}}.
\end{equation}
Taking a sample absorptance of $A=0.02$, the spatially averaged injected exciton density is
\begin{equation}
n_{\mathrm{ex}} \;=\; A\,N_\gamma
\;\approx\; 1.8\times 10^{12}\ \mathrm{cm^{-2}\,pulse^{-1}}.
\end{equation}
This corresponds to an average injected filling of $\langle N_0\rangle \approx 1$ exciton per moiré unit cell
in WSe$_2$/WS$_2$.
%the pulse energy is the integral over the focal plane,
%\begin{equation}
%\label{eq:pulse_energy}
%E \;=\; \iint F(x,y)\,dx\,dy \;=\; %2\pi\sigma^2 F_0,
%\end{equation}
%where $F_0$ is the peak fluence at the beam center and $\sigma$ is the Gaussian width
%(defined by the $1/e$ radius of the fluence; see Methods for beam-size conventions).

%For a repetition rate $f=750~\mathrm{kHz}$ and average power $P=110~\mathrm{nW}$,
%the pulse energy is
%\begin{equation}
%E \;=\; \frac{P}{f} \;=\; 1.47\times %10^{-13}\ \mathrm{J}.
%\end{equation}
%With $\sigma = 0.22~\mu\mathrm{m}$ and the relation \eqref{eq:pulse_energy}, we obtain the peak fluence
%\begin{equation}
%F_0 \;=\; \frac{E}{2\pi\sigma^2}
%\;\approx\; 2.4\times 10^{1}\ %\mu\mathrm{J\,cm^{-2}},
%\end{equation}
%evaluated at the pump photon energy $E_\gamma=1.68~\mathrm{eV}$.
%The corresponding photon areal %density at the beam center is
%\begin{equation}
%N_\gamma \;=\; \frac{F_0}{E_\gamma}
%\;\approx\; %\frac{24~\mu\mathrm{J\,cm^{-2}}}%{1.68~\mathrm{eV}}
%\;\approx\; 9\times 10^{13}\ %\mathrm{cm^{-2}\,pulse^{-1}}.
%\end{equation}
%Taking a sample absorptance of %$A=0.02$, the injected exciton %density at the beam center is
%\begin{equation}
%n_{\mathrm{ex}} \;=\; A\,N_\gamma
%\;\approx\; 1.8\times 10^{12}\ %\mathrm{cm^{-2}\,pulse^{-1}}.
%\end{equation}
%This corresponds to an average %injected filling of $\langle %N_0\rangle \approx 1$ exciton per %moiré unit cell
%in WSe$_2$/WS$_2$.
We independently calibrated the exciton density using power-dependent PL spectroscopy (Extended Data Fig.~1). At each excitation power, the PL spectrum was fitted with two Gaussian components corresponding to the \(X\) and \(XX\) emission bands, and the integrated intensity of each component was extracted. The \(XX\) emission becomes increasingly prominent as double occupation of moiré sites becomes appreciable. We fitted the integrated \(XX\) intensity as a function of excitation power with a continuous piecewise-linear function and assigned the resulting breakpoint to an average injected density of approximately one exciton per moiré unit cell. This spectroscopic calibration is consistent with the fluence-based estimate above and was used to convert excitation power into the initial exciton density \(N_0\).

\noindent\textbf{Determination of doping level and filling factor}

Carrier densities were obtained from a dual--gate parallel--plate capacitor model. The top (bottom) hBN thickness
was measured by atomic force microscopy (AFM), and the corresponding gate capacitances were evaluated as
\begin{equation}
  C_{t(b)}=\frac{\varepsilon_0\,\varepsilon_{\mathrm{hBN}}}{t_{t(b)}}, \qquad \varepsilon_{\mathrm{hBN}}\approx 3,
\end{equation}
where \(t_{t(b)}\) denotes the top (bottom) hBN thickness. For gate voltages referenced to the band edges,
\(\Delta V_t\) and \(\Delta V_b\), the two--dimensional carrier density is
\begin{equation}
  n=\frac{C_t\,\Delta V_t + C_b\,\Delta V_b}{e},
\end{equation}
with \(e\) the elementary charge. The moir\'e lattice constant \(a_M\), determined by piezoresponse force microscopy (PFM),
sets the moir\'e unit--cell density
\begin{equation}
  n_0=\frac{2}{\sqrt{3}\,a_M^2},
\end{equation}
so that the filling factor is
\begin{equation}
  \nu=\frac{n}{n_0}.
\end{equation}
Filling--factor assignments were corroborated by gate--dependent differential reflectance and photoluminescence.\newline

\noindent\textbf{Lifetime and diffusion-constant fitting}

Time-resolved photoluminescence traces were fit with a bi-exponential decay convolved with a Gaussian instrument response. Specifically, the signal is modeled as
\begin{equation}
S(t)=A_{\mathrm{f}}\,e^{-t/\tau_{\mathrm{f}}}+A_{\mathrm{s}}\,e^{-t/\tau_{\mathrm{s}}}, \qquad
G(t)=\frac{1}{\sqrt{2\pi}\,\sigma_{\mathrm{IRF}}}\exp\!\left[-\frac{(t-t_0)^2}{2\sigma_{\mathrm{IRF}}^2}\right],
\end{equation}
and the measured intensity is
\begin{equation}
I(t)=\int_{-\infty}^{\infty} S(t')\,G(t-t')\,dt'.
\end{equation}
Here, \(\tau_{\mathrm{f}}\) and \(\tau_{\mathrm{s}}\) denote the fast and slow decay constants, respectively associated with excitons co-localized with an electron and single-exciton decay. In the global analysis across electron fillings, \(\tau_{\mathrm{f}}\) and \(\tau_{\mathrm{s}}\) are shared for all datasets, while the amplitudes \(A_{\mathrm{f}},A_{\mathrm{s}}\) and background offsets vary per filling. Fits were obtained by nonlinear least squares with the Gaussian width \(\sigma_{\mathrm{IRF}}\).

To extract the time-dependent diffusion constant \(D(t)\), we evaluate the time derivative of the exciton mean-squared displacement (MSD) using equation \eqref{eq:MSD} in the main text. The derivative is computed from binned MSD data using finite differences with a 3~ns bin size for \(t\in[0,30]~\mathrm{ns}\) and a 10~ns bin size for \(t>30~\mathrm{ns}\).\newline

\noindent\textbf{Numerical simulations}\\\noindent
We simulate the non-equilibrium dynamics of the excitons-electron mixture by employing the Keldysh formalism\cite{Stefanucci}\cite{SchlünzenRev}. All interactions are treated on the self-consistent Hartree-Fock mean-field level.
% The electronic density is obtained by solving the equilibrium Hartree-Fock equation in the absence of excitons. 
To model the excitonic system we use an open, extended two-band Bose-Hubbard model 
\begin{equation}\label{eq:exciton_model}
    \begin{aligned}
    \hat{H}^{x} =& 
            \,\Delta\sum_i \hat{n}_{i2} 
            -\sum_{\langle i,j\rangle}\sum_{n}J^{\vphantom{\dagger}}_{n} \left(\hat{b}^{\dagger}_{in} \hat{b}^{\vphantom{\dagger}}_{jn}+\text{h.c.}\right)
            -\sum_{i}J^{\vphantom{\dagger}}_{12} \left(\hat{b}^{\dagger}_{i2} \hat{b}^{\vphantom{\dagger}}_{i1}+\text{h.c.}\right)\\
            &+\frac{U^{xx}}{2}\sum_i\sum_{n\,n'} \hat{n}_{in} \left(\hat{n}_{in'}-\delta_{nn'}\right)
            +\frac{1}{2}\sum_{i \neq i'}V_{ii'}^{xx}\sum_{n\,n'} \hat{n}_{in} \hat{n}_{i'n'}
            +\hat{H}^{xe}
            +\hat{H}^\text{res}
    \end{aligned}
\end{equation}
on a two-dimensional triangular lattice with lattice constant $a$; with intra-band nearest-neighbor hopping $J_n$, on site interactions $U^{xx}$ and long range dipole-dipole interactions $V_{ii'}^{xx}=V^{xx}\frac{a^3}{|\mathbf{r}_i-\mathbf{r}_{i'}|^3}$. Here $\hat{b}_i^\dagger$ ($\hat{b}_i$) denote the creation (annihilation) operators for excitons on site $i$ and $\hat{n}_i = \hat{b}_i^\dagger \hat{b}_i$ are the corresponding number operators. In the notation, $n$ and $n'$ are band indices and $\langle i,j\rangle$ indicates nearest neighbors. $\Delta$ denotes the energy separation of the bands, and $J_{12}$ is the inter-band hopping amplitude.
The term $\hat{H}_{xe}$ represents the interaction of the excitons with the electrons and the term $\hat{H}^\text{res}$ the coupling to and Hamiltonian of bosonic reservoirs. \newline

% In the experiment, an inert lower band and a dispersive upper band were observed. The free parameters in the simulation were chosen accordingly.
% We select $V^{xx}=2J_1$ so that the dispersion of excitons in the lower band is frozen.
% The hopping of the upper band is significantly larger than the hopping of the lower band, $J_2 = 10 J_1$. \newline
% We use the two band model to describe the different energy states observed experimentally for Moiré sites with only one exciton or with an electron and an exciton !!!ref to figure!!!.
% The energy of two excitons occupying the same site is  experimentally observed to be highest. Because we model the excitonic filling sector, where only at most one exciton occupies a site in the beginning and focus on the effect of electron-exciton interactions, we do not model a third band explicitly and set $U_x=J_1$. \newline
\noindent The two-band model \eqref{eq:exciton_model} is employed to describe two of the experimentally observed states, specifically $X$ and $Xe$ as depicted in Fig.~\ref{fig:fig1}. Since we are interested in the situation where only at most one exciton occupies a site and focus on the effect of electron-exciton interactions, we do not model the third $XX$ band explicitly and set $U^{xx}=J_1$.
The other parameters are then chosen to reflect the experimental observations, namely a basically inert lower band $X$ and a dispersive upper band $Xe$. We chose $V^{xx}=1.2J_1$ so that the dispersion of excitons in the lower band is small. On the other hand, the hopping of the upper band is significantly larger than the hopping of the lower band, $J_2 = 10 J_1$. For the energy separation of the bands we use $\Delta=20J_2$ which is much larger than the bandwidth of the upper or lower bands. We set $J_{12}=J_2$. This allows for similar timescales of particle transfer to the upper band and the dynamics of the particles in it. 
The inter-band hopping accounts for transfer between the Moiré-bands caused by scattering.    
Interactions of particles in the lower band with the electronic background charge may increase the potential energy such that the transfer to the upper band becomes energetically more favorable. We simplify the model by using a fixed inter-band hopping $J_{12}$ instead of some scattering term involving electronic operators. We argue that at least in mean-field approximation, including electronic operators would only change the magnitude of hopping, which mainly has an effect on the timescales of particle transfer and less on its magnitude. We include a strong onsite repulsion of electrons and excitons which is modeled with a density-density interaction
\begin{equation*}
    \hat{H}^{xe} = U^{xe} \sum_i \hat{n}_i^e\hat{n}_{i1}
    \,,
\end{equation*}
 with strength $U^{xe}=\Delta$. Here $\hat{n}_i^e:=\left\langle \hat{c}_{i}^\dagger \hat{c}_{i}^{\vphantom{\dagger}} \right\rangle$ is the electronic particle number operator and $\hat{c}^\dagger_i$ ($\hat{c}_i$) denote creation (annihilation) operators for fermions on site $i$.
%The bands explicitly introduced model the two lowest bands caused by the periodic Moiré-potential. 
%One should keep in mind that interactions, with a periodic GWC electron background, for example, can split and deform these bands. 

 %It is worth mentioning that the microscopic form of this correlated hopping is in general the outcome of a  down-folding procedure which gives rise to in general all symmetry allowed terms. Usually most of them are discarded at model-building. Here we focus on the coupling to $ \left\langle \hat{n}^e_i \right\rangle^2$ as one of those terms. We have also tested including first order terms $\propto \left\langle \hat{n}^e_i \right\rangle$ which produced very similar results. This identifies the second order coupling term as relevant for the effects shown here. The exact form of the coupling and whether down-folding produces other significant contributions depends on the excitonic and electronic wavefunctions and is out of scope for this investigation. 
 
\noindent 
Since the interaction is spin-independent and the doublon-state of electrons is irrelevant to the physics we discuss, the electrons are modeled as spinless fermions with nearest neighbor interactions
\begin{equation}\label{eq:electron_model}
    \hat{H}^{e} = -t\sum_{\langle i,i'\rangle}\left(\hat{c}_i^\dagger\hat{c}_{i'}^{\vphantom{\dagger}}+\text{h.c.}\right)+V^{ee}\sum_{\langle i,i'\rangle} \hat{n}_i^e \hat{n}_{i'}^e
    \,.
\end{equation}
We have explicitly checked that including spin does not change the phyiscs reported here. 
Electrons experience the same Moiré potential as the excitons, their hopping is taken similar to the excitonic hopping in the lower band, $t=J_1$. For the nearest-neighbor interaction we chose $V^{ee}=10J_1$.

We assume that the electrons are initially in equilibrium.
To obtain the initial state, the self-consistent Hartree-Fock equation
\begin{equation}\label{eq:eq-sc}
    -i\mathbf{G}^\mathrm{M}_-(0^-) =
    % \frac{\mathbbm{1}}{2}-\frac{1}{2\pi i}\left[\Psi\left(\frac{1}{2}\left[\mathbbm{1}+i\frac{\beta}{\pi}\mathbf{h}^\text{M}_\text{eff}(\boldsymbol{\rho}_e)\right]\right)-\text{h.c.}\right] 
    % \underset{\mathbf{\Gamma}^e
    % \to \mathbf{0}}{\longrightarrow} 
    n_\text{f}\!\left(\mathbf{h}^\text{M}_\text{HF}\left[\mathbf{G}_-^\mathrm{M}\right]\right)
\end{equation}
for the Matsubara Green's function $\mathbf{G}_-^\mathrm{M}$ is solved. Here $n_\text{f}$ is the fermi function and $\mathbf{h}^\text{M}_\text{HF} = \mathbf{h}+\mathbf{\Sigma}^\text{HF}-\mu\mathbbm{1}$. The matrix $\mathbf{h}$ denotes the non-interacting part of the Hamiltonian $\hat{H}^e$ (in first quantization) and $\mathbf{\Sigma}^\text{HF}$ the Hartree-Fock self-energy obtained from approximating the nearest-neighbor interaction. The temperature is fixed at $T=1/\beta=6 J_1$. A sweep over the chemical potential $\mu$ is performed to obtain different electronic fillings $\nu_e$. The transition between GWC- and fluid phase can be identified by a kink in the filling $\nu_e(\mu)$ (extended data Fig.~\ref{fig:theory}~a). \newline
To solve the fixed point equation Eq.~\eqref{eq:eq-sc} we employ Anderson mixing\cite{aacc}. The initial state is found by comparing multiple randomized initial seeds of the self-consistent loop and choosing the one that minimizes the grand-canonical potential. \newline\newline\noindent
The reservoirs in the exciton model \eqref{eq:exciton_model} are introduced to capture the finite lifetime of the excitons. Each lattice site is coupled to an identical empty reservoir that plays the role of a particle-drain. The corresponding term in eq.~\eqref{eq:exciton_model} is given by $\hat{H}^\text{res}=\sum_{in}\hat{H}^\text{res}_{in}$ with
\begin{equation}\label{eq:b_reservoir}
    \hat{H}^{\mathrm{res}}_{in} = 
\sum_k \epsilon^{\vphantom{\dagger}}_k \hat{d}^{\dagger}_{k} \hat{d}^{\vphantom{\dagger}}_{k}
+\sum_k \left(\gamma^{\vphantom{\dagger}}_{n}\hat{b}^{\dagger}_{in}\hat{d}^{\vphantom{\dagger}}_k+\text{h.c.}\right)
\,,
\end{equation}
and the creation (annihilation) operators $\hat{d}_k^\dagger$ ($\hat{d}_k$) of the reservoir.
The thermodynamic limit is performed assuming a flat spectral density $D$ (wide-band limit) for the reservoirs. This makes the reservoirs Markovian. They can then be integrated out exactly leading to a dressing of the system's Green's functions via a self-energy contribution $\mathbf{\Sigma}^\text{R}_\text{emb}(t,t')=-i\mathbf{\Gamma}\delta(t-t')$, where $\Gamma_{in\,i'n'}=\delta_{ii'}\delta_{nn'}\pi D |\gamma_n|^2$ \cite{KJKMPaper}. Motivated by the higher decay rate of the upper band observed in experiment, the reservoir coupling $\gamma_2$ is chosen slightly higher than $\gamma_1$, so that $\Gamma_{i2\,i2}=3\Gamma_{i1\,i1}=0.3 J_2$.\\

\noindent Finally, the non-equilibrium dynamics of the exciton-electron mixture are modeled using the Keldysh formalism\cite{Stefanucci}\cite{SchlünzenRev},
where the single particle density matrices are obtained via the lesser Green's functions $\langle \hat{b}^\dagger_{i'n'} \hat{b}^{\vphantom{\dagger}}_{in}\rangle(t) = i\left[\mathbf{G}_+^<(t,t)\right]_{in\,i'n'}$ and $\langle \hat{c}^\dagger_{i'} \hat{c}^{\vphantom{\dagger}}_{i}\rangle(t) = -i\left[\mathbf{G}_-^<(t,t)\right]_{i\,i'}$, which are evolved via the Kadanoff-Baym equations \cite{SchlünzenRev}
\begin{equation}\label{eq:KBE}
    i\frac{d}{dt}\mathbf{G}_\pm^<(t,t) 
    = \mathbf{h}_\pm(t)\mathbf{G}_\pm^<(t,t)
    +\int_{-\infty}^\infty \!ds \left[\mathbf{\Sigma}_\pm^\text{R}(t,s)\mathbf{G}_\pm^<(s,t)
    +\mathbf{\Sigma}_\pm^<(t,s)\mathbf{G}_\pm^\text{A}(s,t')\right]
    +\text{h.c.}
\end{equation}
with the advanced Green's functions $\mathbf{G}_\pm^\text{A}(t,t')$. Note that we use the index $\pm$ to differentiate between excitonic (bosons) and electronic (fermions) quantities.
The single particle Hamiltonians $\mathbf{h}_\pm(t)$ are again obtained from eq.~\eqref{eq:exciton_model} and eq.~\eqref{eq:electron_model} respectively, by the non-interacting terms of the system part (not $\hat{H}^\text{res})$ in first quantization. 
Inserting the self-energies $\mathbf{\Sigma}_+^\text{R}(t,t')=\delta(t-t')\mathbf{\Sigma}_+^\text{HF}(t)+\mathbf{\Sigma}^\text{R}_\text{emb}(t,t')$, $\mathbf{\Sigma}_-^\text{R}(t,t')=\delta(t-t')\mathbf{\Sigma}_-^\text{HF}(t)$ and $\mathbf{\Sigma}_\pm^<(t,t')=\mathbf{0}$ simplifies~\eqref{eq:KBE} to a coupled system of ordinary differential equations
\begin{equation*}
    \begin{aligned}
    i\frac{d}{dt}\mathbf{G}_+^<(t,t) 
    &= \left[\mathbf{h}_+(t)+\mathbf{\Sigma}_+^\text{HF}(t)-i\mathbf{\Gamma}\right]\mathbf{G}_+^<(t,t)+\text{h.c.}\,,\\
    i\frac{d}{dt}\mathbf{G}_-^<(t,t) 
    &= \left[\mathbf{h}_-(t)+\mathbf{\Sigma}_-^\text{HF}(t)\right]\mathbf{G}_-^<(t,t)+\text{h.c.}\,,\\
    \end{aligned}
\end{equation*}
that is solved via adaptive Runge–Kutta Dormand–Prince method from the Boost.Odeint library\cite{boost:odeint}. The Hartree–Fock self-energies $\mathbf{\Sigma}_\pm^{\text{HF}}(t)$ quantify the mean field experienced by a particle due to its interaction with other particles and thus depends on both $\mathbf{G}_+^<(t,t)$ and $\mathbf{G}_-^<(t,t)$. The explicit expression can be found, for example, in the review by Schlünzen et al.~\cite{SchlünzenRev}.\\
For the excitons, we start from an uncorrelated single particle density, where $60$ sites in a circular center region are singly occupied by excitons in the lower band, and let the excitonic cloud expand into the lattice on top of the electronic background. For the electrons, we use the equilibrium initial state as detailed earlier, with $\mathbf{G}_-^<(t_0,t_0)=\mathbf{G}^\mathrm{M}_-(0^-)$. The lattice has $30\times30$ sites and open boundary conditions are used. To ensure that no considerable effect from particles reaching the simulation boundary is present, the MSD and occupations are monitored and the simulation time and size chosen accordingly. Snapshots of the exciton occupations in the upper band for two exemplary time evolutions can be found in the extended data Fig.~\ref{fig:theory}~b.\\
Our observable, the MSD, is computed via 
\begin{equation}
    \begin{aligned}
        \text{MSD}(t)&=R^2(t)-R^2(t_0) \,,\\
        R^2(t)&=\frac{1}{N(t)}\sum_{in} \langle \hat{n}_{in}\rangle(t) ||\textbf{r}_i-\textbf{s}_0||^2\,,\\
        \textbf{s}_0 &= \frac{1}{N(t_0)}\sum_{in} \langle \hat{n}_{in}\rangle(t_0)\,\textbf{r}_i\,,\\
        N(t) &= \sum_{in} \langle \hat{n}_{in}\rangle(t)\,.
    \end{aligned}
\end{equation}

\subsection*{Phenomenological Coupled Transport--Relaxation Model}

To describe the coupled evolution of exciton transport, interband relaxation, and population decay, we employ a phenomenological coupled transport--relaxation model consisting of a mobile upper-energy exciton state and a lower-energy exciton state. Following photoexcitation and relaxation from higher-energy states, excitons populate the upper state, where they undergo ballistic radial transport while simultaneously relaxing into the lower state. In addition to this transfer, the upper-state population may also be depleted through competing loss processes that do not populate the lower state. The lower state exhibits slow diffusive transport together with population decay. The coupled evolution of the upper- and lower-state populations is described by

\begin{gather}
\frac{\partial u}{\partial t}
=
f(v)P(r,t)
-
k_UA(L)u
-
k_{\mathrm{loss}}u
-
\frac{1}{r}
\frac{\partial}{\partial r}
\left(
rvu
\right),
\\[1em]
\frac{\partial L}{\partial t}
=
k_UA(L)U
-
\gamma_LL
+
D_L
\frac{1}{r}
\frac{\partial}{\partial r}
\left(
r\frac{\partial L}{\partial r}
\right).
\end{gather}

Ballistic transport in the upper state is described by a distribution of outward radial velocities. The upper-state population $u(r,v,t)$ is treated as a radially symmetric areal density per unit velocity, while $L(r,t)$ is the lower-state areal density. The total upper-state population is given by

\begin{equation}
U(r,t)
=
\int_0^\infty u(r,v,t)\,dv.
\end{equation}

The injected excitons have a normalized distribution of positive velocities given by

\begin{equation}
f(v)
=
\frac{1}{\mathcal N}
\exp\!\left[
-\frac{(v-v_g)^2}{2\sigma_v^2}
\right],
\qquad
v>0,
\end{equation}

where $\mathcal N$ is the normalization constant satisfying

\begin{equation}
\int_0^\infty f(v)\,dv=1.
\end{equation}

Here, $v_g$ and $\sigma_v$ are the center and width parameters of the underlying Gaussian before truncation to positive velocities. The velocity parameter $v_g$ is extracted from the fit. For numerical implementation, the continuous velocity distribution is discretized into a finite number of velocity channels.

Photoexcitation generates population in the upper state through a Gaussian spatiotemporal pump profile,

\begin{equation}
P(r,t)
=
P_0
\exp\!\left(-\frac{r^2}{2\sigma_r^2}\right)
\exp\!\left(-\frac{(t-t_0)^2}{2\sigma_t^2}\right).
\end{equation}

This source term assumes that relaxation into the upper state is fast compared with the experimental time resolution.

Transfer from the upper to the lower state occurs at a rate proportional to $k_UA(L)$, where

\begin{equation}
A(L)
=
\frac{1}{1+\left(\frac{L}{L_{\max}}\right)^n}.
\end{equation}

The function $A(L)$ represents an occupancy-dependent suppression of exciton capture. As the lower state population increases, the availability of lower state decreases, reducing the transfer rate from the upper state. The parameter $L_{\max}$ is the density at which capture is reduced by half, rather than a strict upper bound on the population. The parameter $k_U$ describes interband relaxation from the upper state into the lower emissive state, while $k_{\mathrm{loss}}$ represents competing first-order upper-state loss processes that do not populate the lower state.

The upper-state population propagates ballistically through the radial flux-divergence term $\frac{1}{r}\frac{\partial}{\partial r}\left(rvu\right)$, which accounts for the geometric dilution of an outward-moving areal density. The lower energy population evolves through radial diffusion and population decay according to the diffusion term $D_L\frac{1}{r}\frac{\partial}{\partial r}\left(r\frac{\partial L}{\partial r}\right)$ and the decay term $-\gamma_LL$.

Because the measured photoluminescence at a given emission energy is determined by the combined effects of transport, interband relaxation, and population dynamics, the model is simultaneously fitted to the spatially and temporally resolved photoluminescence measured at multiple emission energies, including the XX and X emission. Simultaneously fitting these coupled observables enables transport, interband relaxation, and population dynamics to be determined self-consistently.

To reduce parameter degeneracy, several experimental parameters were fixed during the fitting procedure. The spatial and temporal widths of the excitation pulse were fixed at 400~nm and 300~fs, respectively, based on the measured optical response of the experimental system. The excitation density was specified using the experimentally determined spatially averaged Gaussian exciton density, expressed relative to the saturation threshold, consistent with the procedure used throughout the experimental analysis. The saturation exponent $n$ was fixed at 6 to represent strong occupancy-dependent suppression of capture.

The term $k_{\mathrm{loss}}$ describes population lost from the upper state without populating the lower emissive state. Its magnitude relative to $k_U$ was constrained by fitting the spectrally, spatially, and temporally integrated photoluminescence as a function of excitation power. Integration accounts for redistribution of emission in energy, space, and time, allowing the total photoluminescence yield to constrain competing population loss within the model (Extended Data Fig.~10).

Finally, to maintain model simplicity, we compare two representative velocity distributions: (i) a finite-width truncated Gaussian distribution whose width parameter $\sigma_v$ is comparable to its center parameter $v_g$ and (ii) the limiting case of zero velocity variance. The finite-width distribution consistently provides a better description of the measured spatiotemporal evolution, yielding significantly lower residual error. Consequently, this distribution is adopted for all subsequent fitting, while the comparison between the two cases is presented in Extended Data Fig.~6.

\section*{Acknowledgments} This work is mainly supported by U.S. Department of Energy, Office of Basic Energy Sciences through QuPIDC EFRC award DE-SC0025620 (for the transient and time-resolved optical spectroscopy and microscopy work at Purdue). We also acknowledge funding from the Deutsche Forschungsgemeinschaft (DFG, German Research Foundation) - 531215165 (Research Unit ``OPTIMAL'') for the theory contributions. This work was partially supported by the Max Planck-New York City Center for Non-Equilibrium Quantum Phenomena. The Flatiron Institute is a division of the Simons Foundation. The work at U. Washington is supported by DoE, BES under the award DE-SC0018171.  K.W. and T.T. acknowledge support from the JSPS KAKENHI (Grant Numbers 21H05233 and 23H02052), the CREST (JPMJCR24A5), JST and World Premier International Research Center Initiative (WPI), MEXT, Japan.
\section*{Author Contributions}
L.H.and D.M.K. designed the experiments and theoretical simulations; S.D. and J.P. carried out the optical measurements; H.P. and X.X. fabricated and characterized samples; M.H., J.R., A.F. and D.M.K. carried out and analyzed the time-dependent mean-field calculations; K.W. and T.T. grew the hBN crystals; L.H. wrote the manuscript with input from all authors.
 
\section*{Data availability}
All data supporting the findings of this study are available within the paper and its Extended Data. Source data are provided with this paper.
\section*{Code availability}
Codes that support the findings of this study are available upon request. Codes include scripts for data processing and theoretical modeling.  

\printbibliography

\newpage

\section*{Extended Data}
\FloatBarrier
\captionsetup[figure]{name=Extended Data Figure}
\setcounter{figure}{0}

\begin{figure}[H]
    \centering
    \includegraphics[width=1\linewidth]{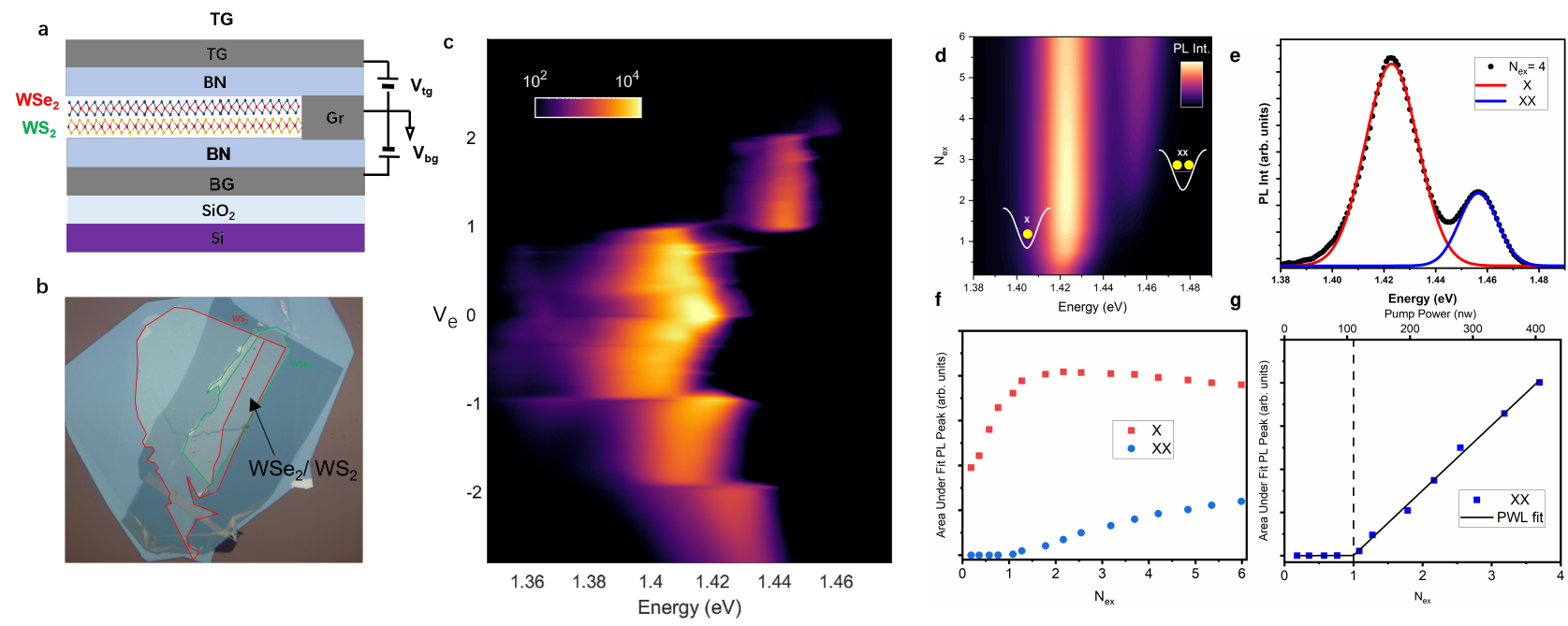}
    \caption{Device configuration, optical characterization, and determination of the injected exciton density. \textbf{a,} Schematic of the dual-gated, H-stacked WSe$_2$/WS$_2$ moiré heterobilayer encapsulated by hBN, with graphite gates and Pt contacts. \textbf{b,} Optical image of the completed device used for time- and energy-resolved photoluminescence (PL) measurements at 6~K. \textbf{c,} Gate-dependent PL spectra of the device under continuous-wave excitation with a 740~nm laser. \textbf{d,} Power-dependent PL spectra showing the emergence of the higher-energy XX emission associated with doubly occupied moiré sites. \textbf{e,} Representative high-power spectrum and double-Gaussian fit resolving the X and XX emission peaks. \textbf{f,} Integrated areas of the X and XX emission peaks, extracted from Gaussian fits, as a function of excitation power. \textbf{g,} Integrated XX PL intensity as a function of pump power (top axis), fitted by a piecewise-linear function. The resulting inflection point identifies the excitation density corresponding to approximately one injected exciton per moiré site.
}
    \label{fig:ed1}
\end{figure}
\newpage
%\begin{figure}[H]
%    \centering
%    \includegraphics[width=1\linewidth]{extd fig 6.png}
%    \caption{Determination of the injected exciton density. \textbf{a,} Power-dependent photoluminescence spectra showing the emergence of the higher-energy XX emission associated with doubly occupied moiré sites. \textbf{b,} Integrated areas of the X and XX emission peaks, extracted from Gaussian fits, as a function of excitation power. \textbf{c,} Representative high-power spectrum and double-Gaussian fit resolving the X and XX emission peaks. \textbf{d,} Integrated XX photoluminescence intensity as a function of pump power (top axis), fitted by a piecewise-linear function. The resulting inflection point identifies the excitation density corresponding to approximately one injected exciton per moiré site.}
%    \label{fig:placeholder}
%\end{figure}
\begin{figure}[H]
    \centering
    \includegraphics[width=1\linewidth]{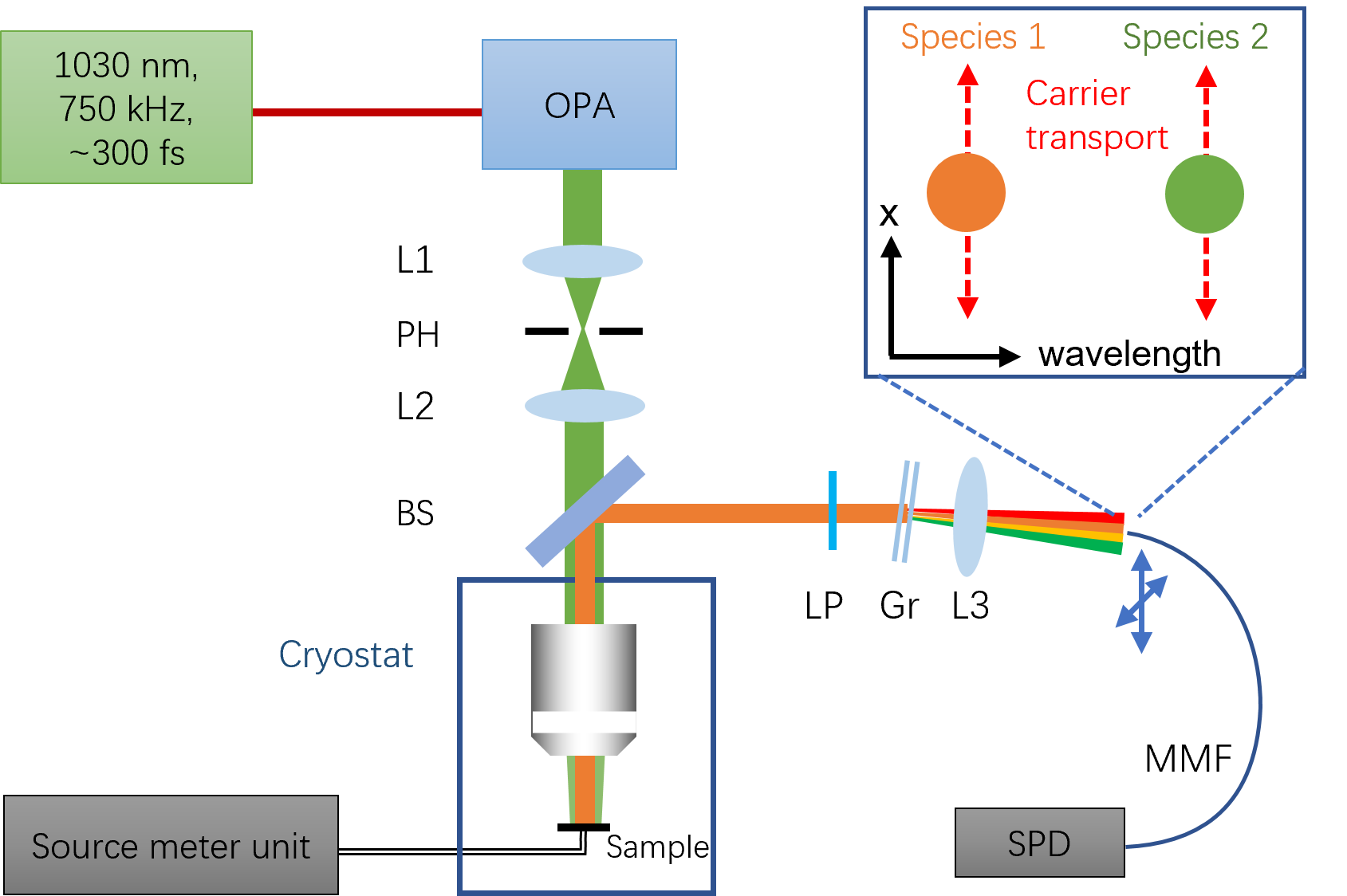}
    \caption{Spectrally resolved transient PL imaging setup. OPA-pumped, diffraction-limited excitation and epi-collection geometry with spectral dispersion prior to fiber coupling for raster-scanned, single-photon counting detection. Abbreviations: OPA, optical parametric amplifier; L, lens; PH, pinhole; BS, beam splitter; LP, long-pass filter; Gr, grating; MMF, multimode fiber; SPD, single-photon detector.}
    \label{fig:extended-data-4}
\end{figure}
\newpage

\begin{figure}[H]
    \centering
    \includegraphics[width=1\linewidth]{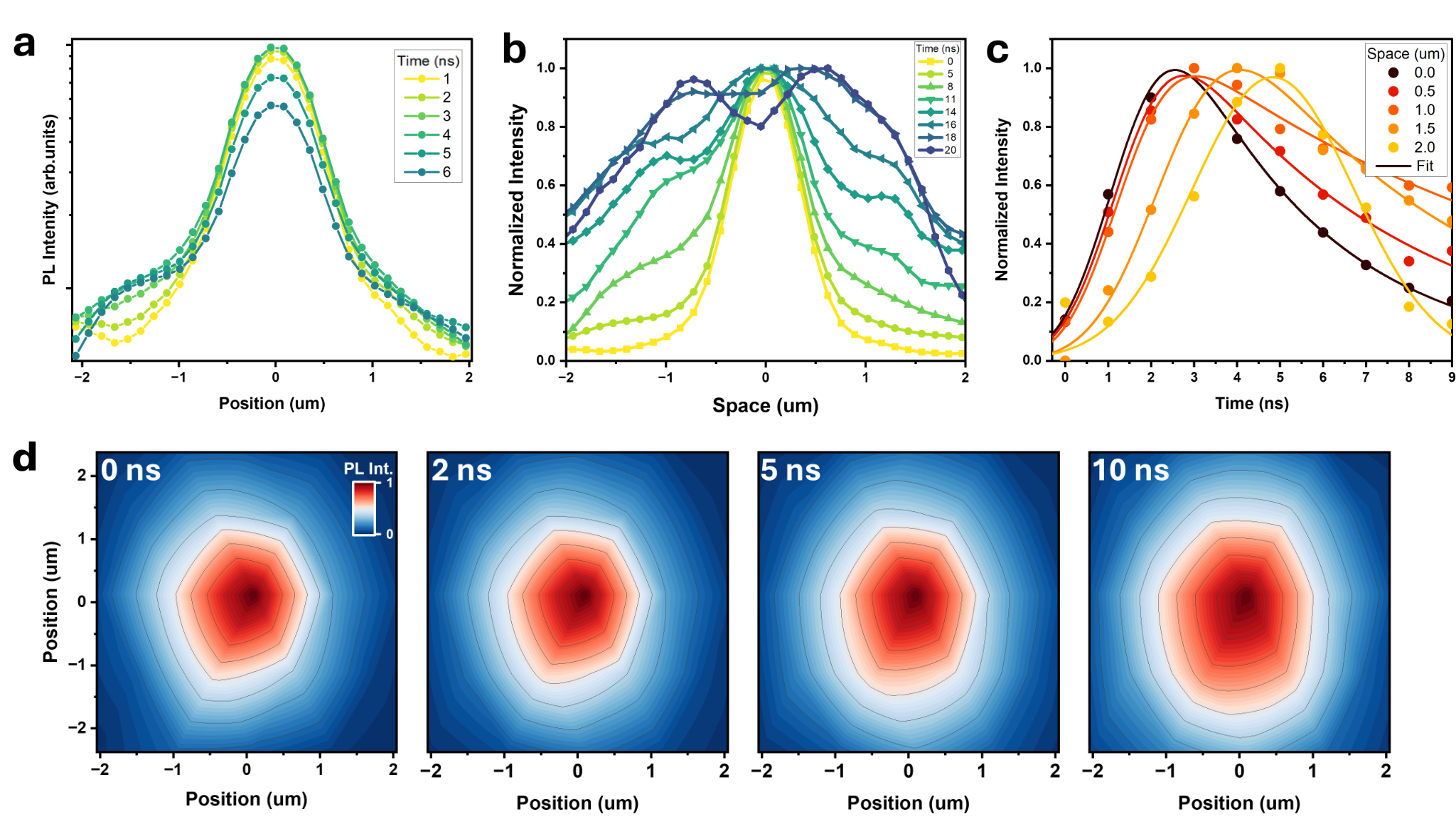}
    \caption{Signatures of ballistic transport in the XX band. \textbf{a,} Unnormalized spatial photoluminescence profiles on a logarithmic scale during the first 6 ns after excitation. The depletion of the central population is accompanied by a rise in the population at the edges of the spatial profile, consistent with outward propagation. \textbf{b,} Spatial profiles normalized to their maximum intensity and displayed on a linear scale. At later times, most clearly near 10 ns, the population separates into two outward-propagating peaks. \textbf{c,} Delay in the photoluminescence rise time as a function of distance from the excitation center. The systematic increase in arrival time with distance is a hallmark of ballistic propagation. \textbf{d,} Two-dimensional evolution of the exciton population. Spectrally integrated two-dimensional photoluminescence intensity at representative time delays (0, 2, 5, and 10 ns).}
    \label{fig:placeholder}
\end{figure}
\begin{figure}[H]
    \centering
    \includegraphics[width=1\linewidth]{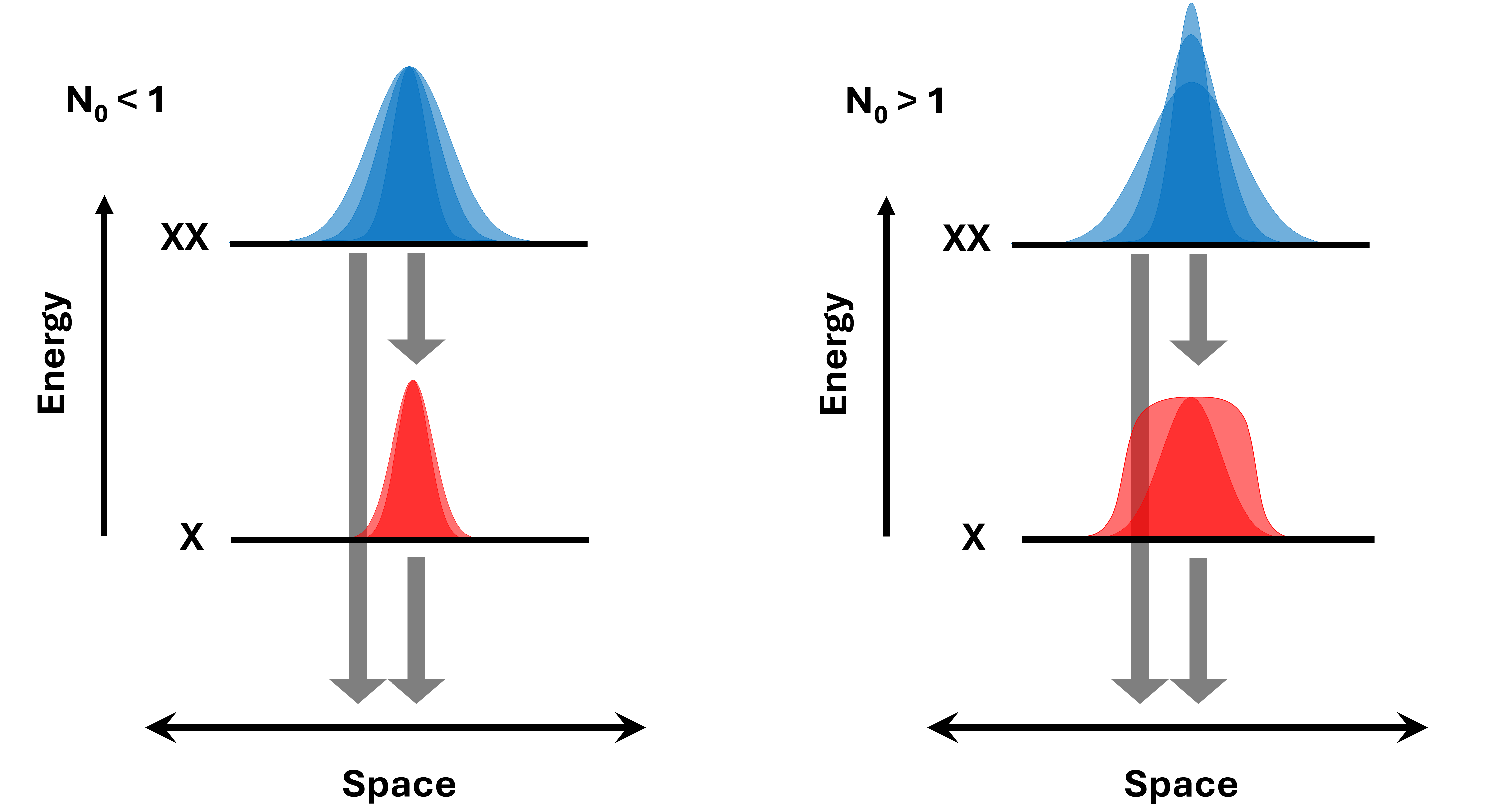}
    \caption{Schematic illustration of phenomenological coupled transport--relaxation model used to extract band-selective transport and decay parameters. \textbf{a,} At low excitation density ($N_0<1$), the higher-energy XX or Xe population undergoes fast ballistic propagation and decay while relaxing into the lower-energy X band, which exhibits slow diffusion and decay. \textbf{b,} At high density ($N_0\gtrsim1$), the X-band occupation approaches saturation near one exciton per moiré site, suppressing further relaxation from the higher-energy band. This occupation-dependent relaxation, together with density-dependent loss and transport in both bands, couples their spatiotemporal evolution and is included when extracting the higher-band group velocity and decay rate.}
    \label{fig:transport-model-schematic}
\end{figure}
\begin{figure}[H]
    \centering
    \includegraphics[width=1\linewidth]{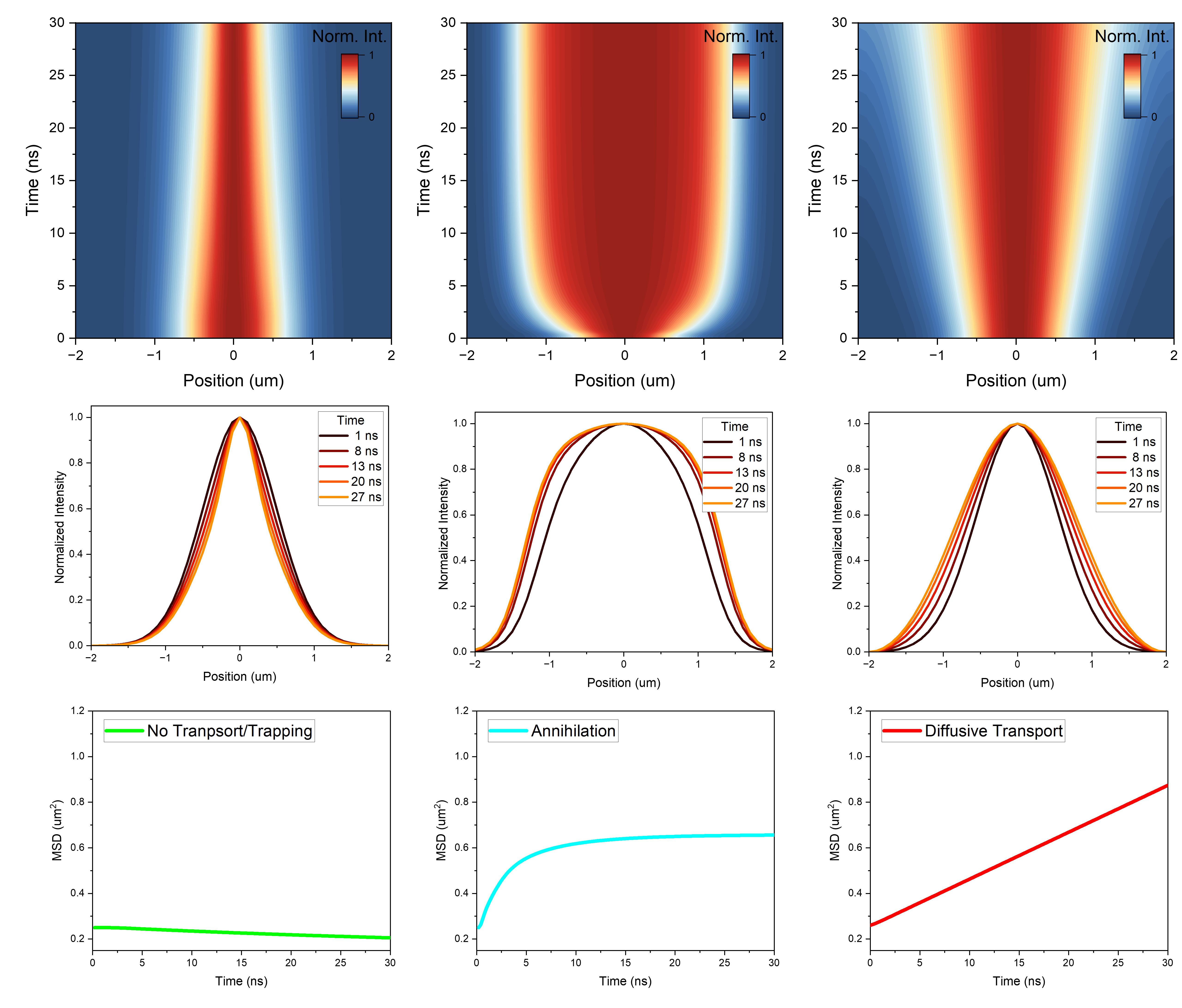}
    \caption{Simulated upper-band evolution for negligible transport,density-dependent annihilation without transport, and diffusive transport.The left column shows negligible transport, the middle column shows density-dependent annihilation without transport, and the right column shows diffusive transport. The top row shows normalized spatial intensity as a function of position and time, the middle row shows representative spatial profiles at selected times, and the bottom row shows the corresponding mean-squared displacement (MSD, $\sigma_x^2$). With negligible transport,the spatial profile remains nearly unchanged. Annihilation preferentially depletes the population near the excitation center, causing apparent broadening and an MSD that rises before saturating. Diffusive transport broadens the profile and produces an approximately linear increase in MSD with time.}
    \label{fig:upper-band-transport-comparison}
\end{figure}

\begin{figure}[H]
    \centering
    \includegraphics[width=0.93\linewidth]{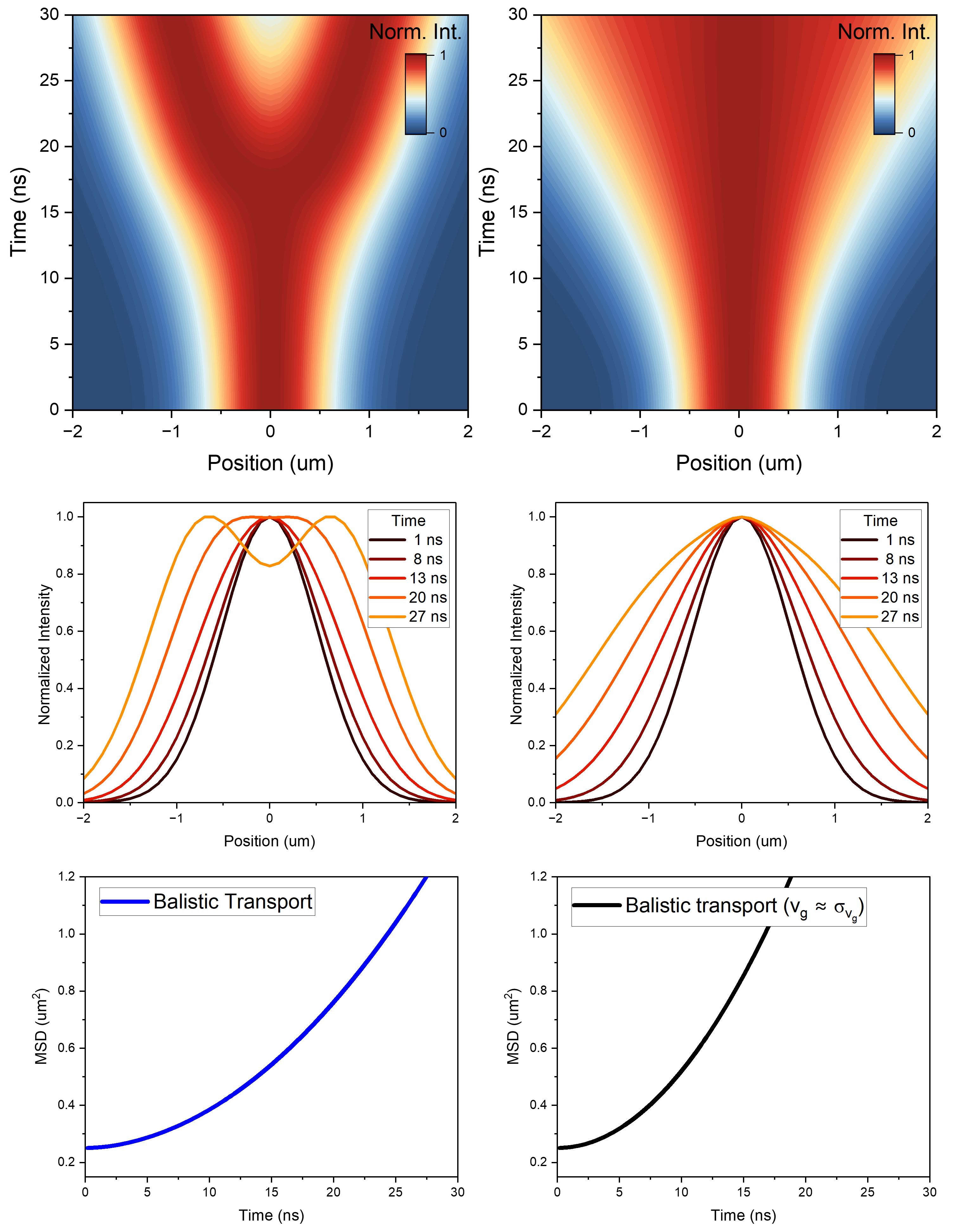}
    \caption{Simulated upper-band evolution for ballistic transport with a single velocity and with a distribution of velocities. The left column shows ballistic transport with a single velocity ($\sigma_v = 0$), which separates the population into two outward-propagating fronts. The right column shows ballistic transport with a broad velocity distribution ($\sigma_v \approx v_g$), which broadens the spatial profile without clearly resolved peak splitting. The top row shows normalized spatial intensity as a function of position and time, the middle row shows representative spatial profiles at selected times, and the bottom row shows the corresponding mean-squared displacement (MSD, $\sigma_x^2$). Both ballistic cases produce an approximately quadratic increase in MSD with time.}
    \label{fig:ballistic-versus-annihilation}
\end{figure}

\begin{figure}[H]
    \centering
    \includegraphics[width=1\linewidth]{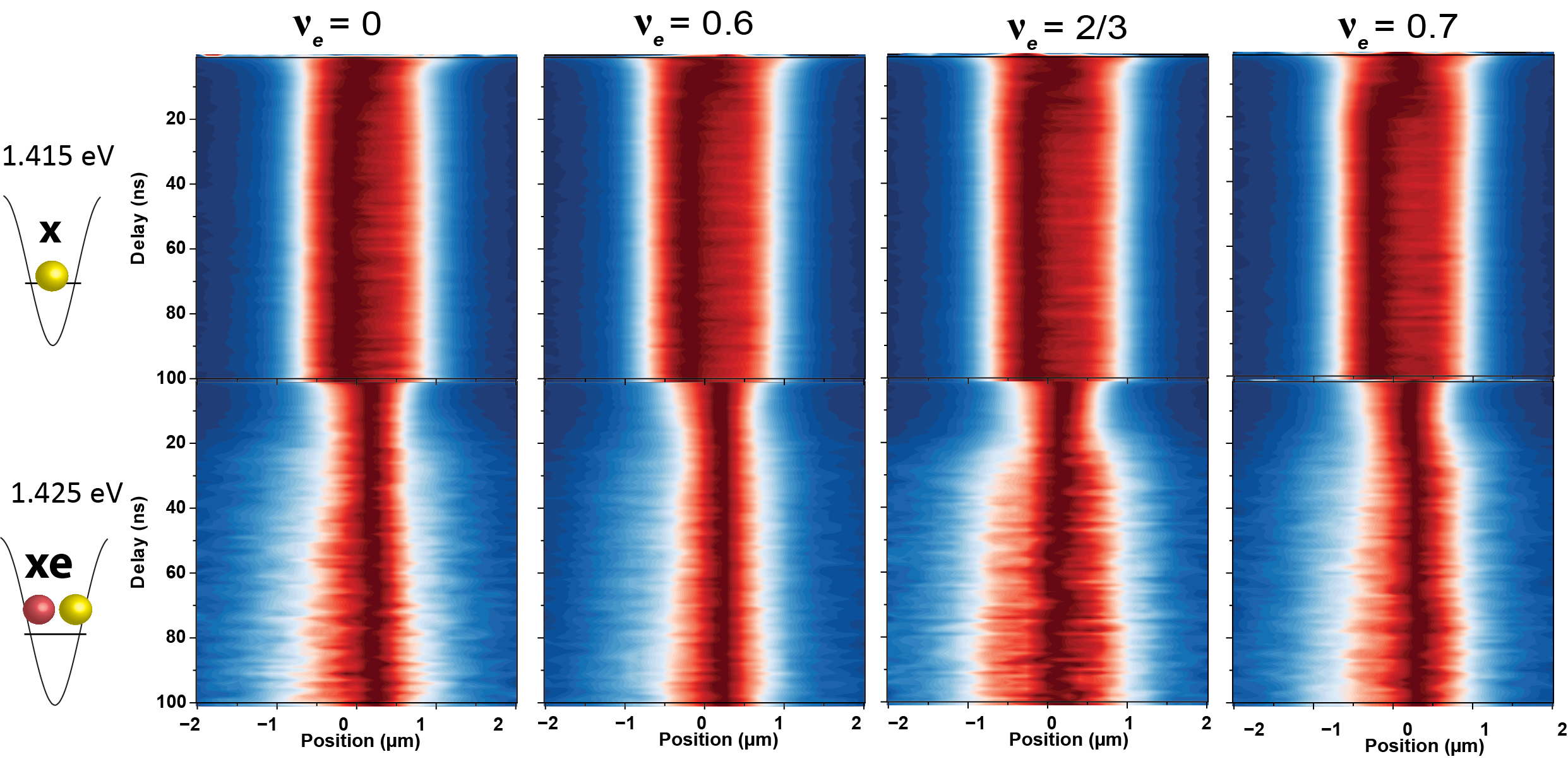}
    \caption{Energy and time-resolved PL images at high exciton density (Mott regime, $N_0 \approx 5$). Representative spatial PL maps versus delay and $\nu_e$ for the blue-shifted (co-localized) exciton emission, showing an enlarged spatial spread and faster decay near GWC fillings (e.g., $\nu_e = 2/3$) that mark accelerated melting of the exciton Mott state.}
    \label{fig:extended-data-5}
\end{figure}
\newpage

\begin{figure}[H]
    \centering
    \includegraphics[width=1\linewidth]{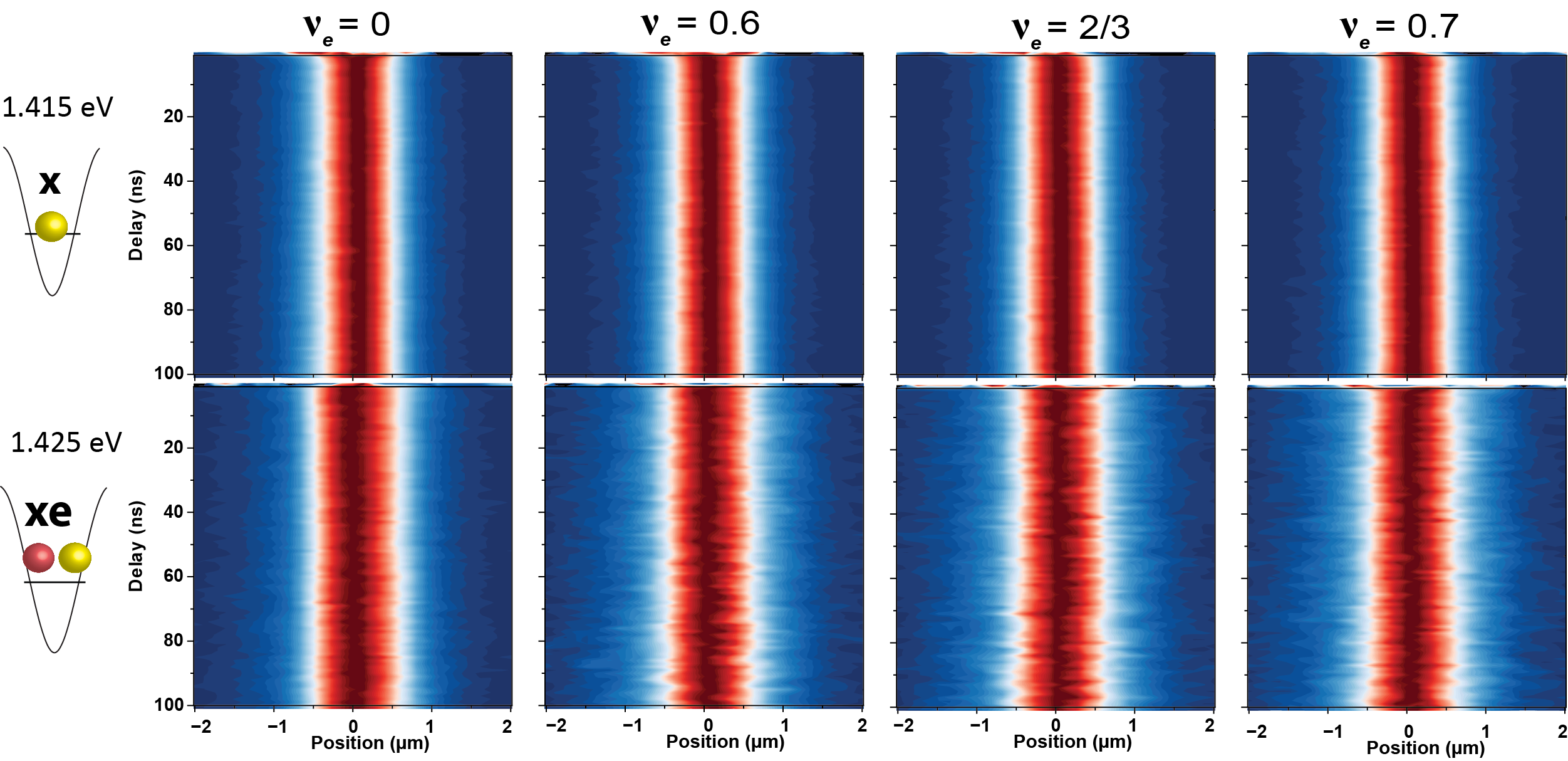}
   \caption{Energy- and time-resolved PL images at low exciton density (weakly correlated, $N_0 \approx 0.3$). Spatial PL evolution for the same spectral window as in Extended Data Fig.~7, revealing a smooth, monotonic increase of transport with $\nu_e$ and the absence of abrupt commensurability-locked changes characteristic of Mott melting.}
    \label{fig:extended-data-6}
\end{figure}
\newpage

%\begin{figure}[H]
%    \centering
%    \includegraphics[width=1\linewidth]{extd fig 6.png}
%    \caption{Determination of the injected exciton density. \textbf{a,} Power-dependent photoluminescence spectra showing the emergence of the higher-energy XX emission associated with doubly occupied moiré sites. \textbf{b,} Integrated areas of the X and XX emission peaks, extracted from Gaussian fits, as a function of excitation power. \textbf{c,} Representative high-power spectrum and double-Gaussian fit resolving the X and XX emission peaks. \textbf{d,} Integrated XX photoluminescence intensity as a function of pump power (top axis), fitted by a piecewise-linear function. The resulting inflection point identifies the excitation density corresponding to approximately one injected exciton per moiré site.}
%    \label{fig:placeholder}
%\end{figure}
\begin{figure}[H]
    \centering
    \includegraphics[width=0.9\linewidth]{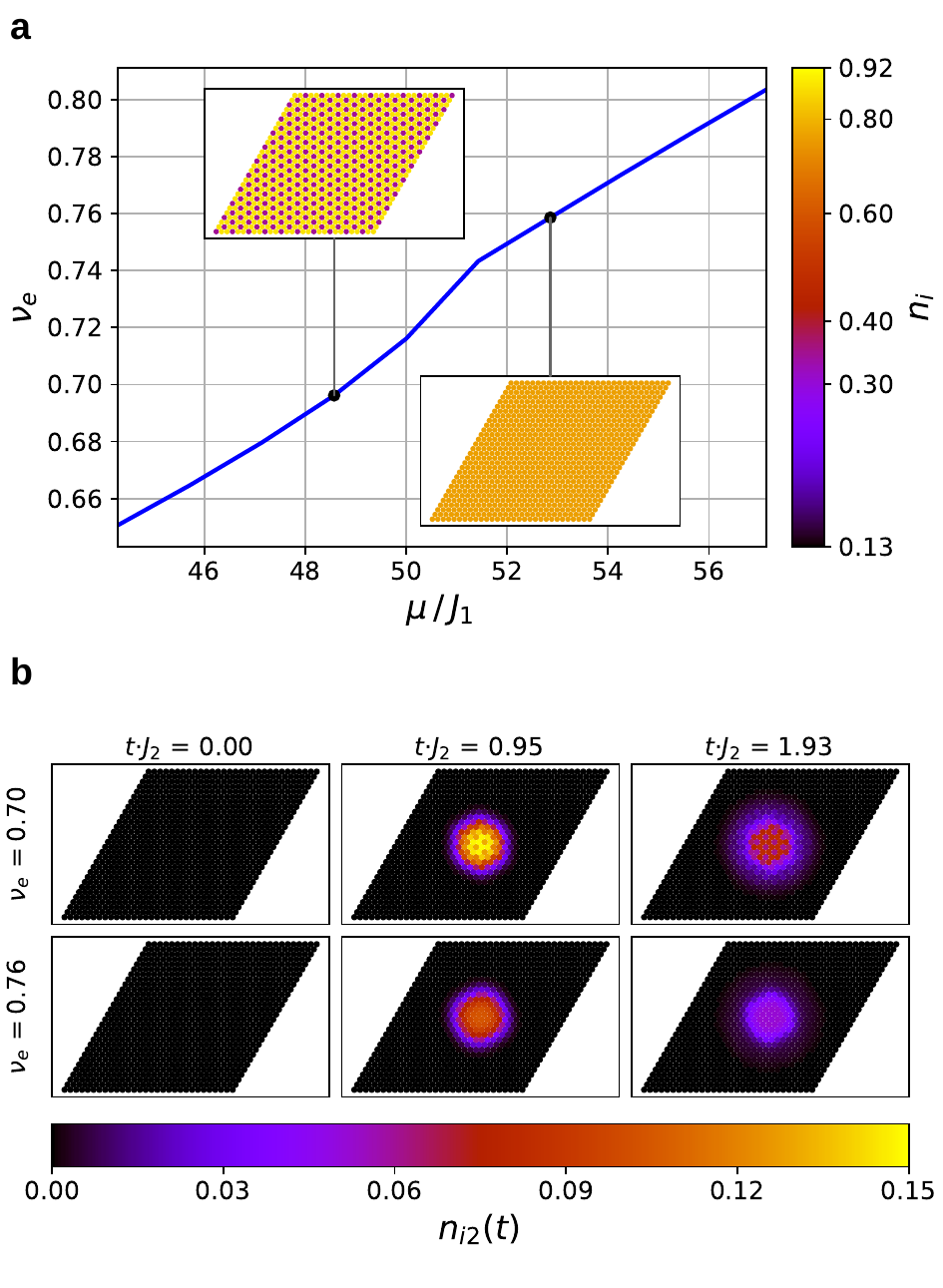}
    \caption{Simulations of a Two-Dimensional Bose-Fermi Mixture. \textbf{a,} Electronic filling $\nu_e$ in dependence of the chemical potential $\mu$ for the solution of the self-consistent Hartree-Fock equation for the electronic initial states. A kink is evident at the transition point between the GWC- and fluid phase. The insets show the electronic occupations $\left\langle\hat{n}^e_i\right\rangle$.
    \textbf{b,} Time evolution of the exciton occupations in the upper band $\left\langle\hat{n}_{i2}\right\rangle$. The electronic background corresponds the two densities displayed in \textbf{a} (upper row GWC- and lower row fluid phase). The exciton population in the upper band is larger when the electrons form a GWC.
    }
    \label{fig:theory}
\end{figure}
\begin{figure}[H]
    \centering
    \includegraphics[width=1\linewidth]{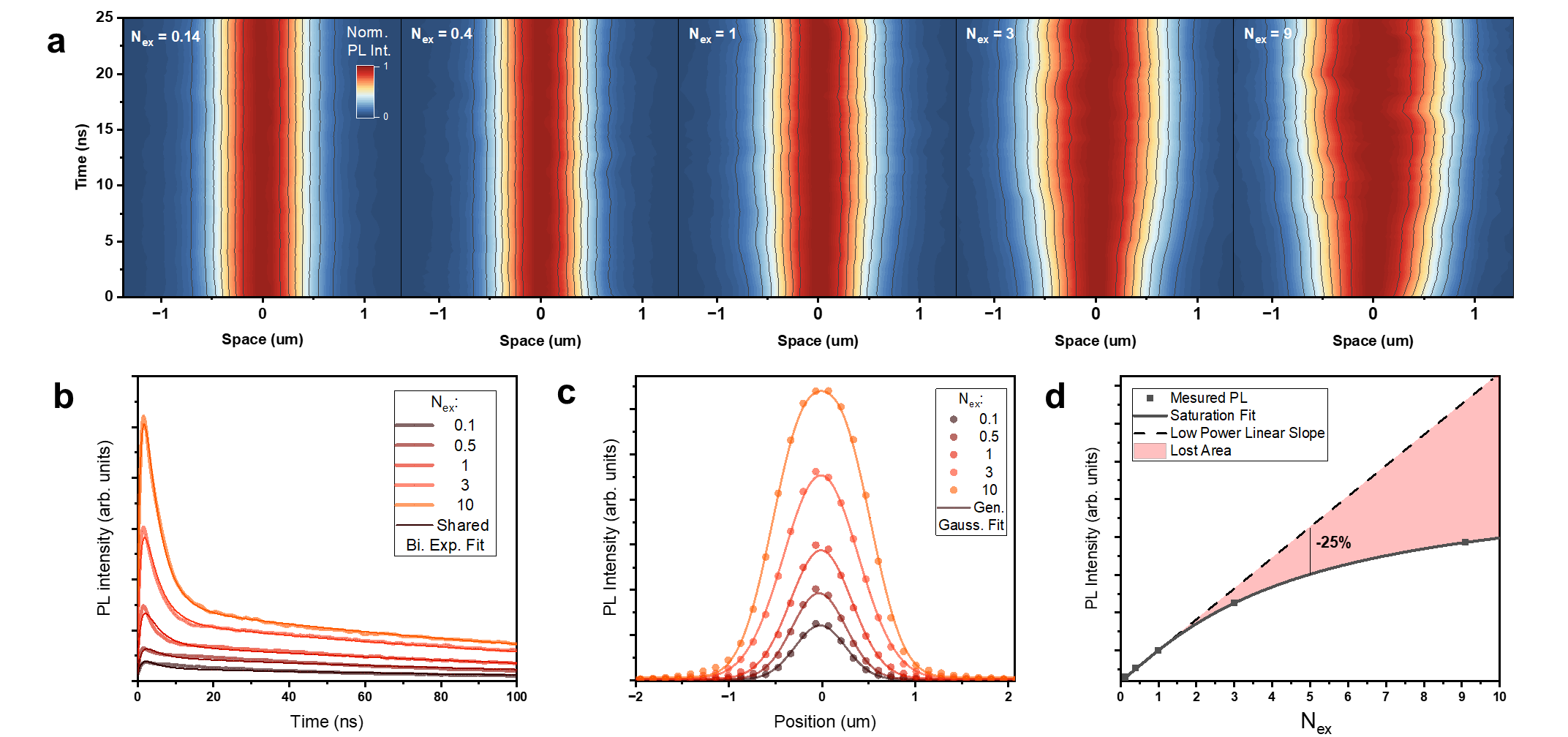}
    \caption{Emergence of enhanced exciton transport above the excitonic Mott transition and determination of exciton population loss. \textbf{a,} Spectrally integrated, normalized spatiotemporal photoluminescence profiles for increasing initial exciton density ($N_{\mathrm{ex}}=0.14$, 0.4, 1, 3, and 9; left to right). Below the excitonic Mott transition, the spatial profile remains relatively localized. Near $N_{\mathrm{ex}}\approx1$, enhanced spatial expansion emerges and becomes progressively more pronounced at higher densities. \textbf{b,} Spectrally and spatially integrated photoluminescence decay as a function of time for increasing excitation density. Above the Mott transition, a pronounced fast-decaying component emerges. \textbf{c,} Corresponding unnormalized spatial photoluminescence profiles integrated over the first 30 ns, together with generalized Gaussian fits. \textbf{d,} Total photoluminescence integrated over energy, space, and time as a function of excitation density. The reduction in integrated photoluminescence quantifies population loss through exciton--exciton annihilation and Auger recombination, which is used to constrain population loss in the transport model.}
    \label{fig:extended-data-2}
\end{figure}
\end{document}